\documentclass[aps,onecolumn,showpacs,nofootinbib]{revtex4}
\usepackage{graphicx}
\usepackage{epsfig}
\usepackage{amssymb}
\usepackage{graphicx}
\usepackage{amsfonts}
\usepackage{latexsym}
\usepackage{amsmath}
\usepackage{subfigure}
\usepackage{mathrsfs}
\numberwithin{equation}{section}

\newcommand\encadremath[1]{\vbox{\hrule\hbox{\vrule\kern8pt
\vbox{\kern8pt \hbox{$\displaystyle #1$}\kern8pt}
\kern8pt\vrule}\hrule}}
\def\enca#1{\vbox{\hrule\hbox{
\vrule\kern8pt\vbox{\kern8pt \hbox{$\displaystyle #1$} \kern8pt}
\kern8pt\vrule}\hrule}}

\newcommand{\tpsi}
{\widehat\Psi}
\newcommand{\tphi}
{\widehat\Phi}

\newcommand{\bt}{(\beta)}
\def\prl#1#2#3{{ Phys. Rev. Lett.} {\bf #1}, #2 (#3)}

\def\pre#1#2#3{Phys. Rev. E  {#1} #2 #3}
\def\jpa#1#2#3{J. Phys. A {\bf #1}, #2 (#3)}

\def\cmp#1#2#3{Commun. Math. Phys. {\bf #1}, #2 (#3)}
\begin{document}
\title{Generalized Christoffel-Darboux formula for classical skew-orthogonal polynomials}
\author{Saugata Ghosh}\email{saugata135@yahoo.com}
\affiliation{233 Green Park, Lake-Town, Kolkata-700055, India.}
\date{\today}
\begin{abstract}
We show that skew-orthogonal functions, defined with respect to
Jacobi weight $w_{a,b}(x)={(1-x)}^a{(1+x)}^b$, $a$, $b>-1$,
including the limiting cases of Laguerre ($w_{a}(x)=x^{a}e^{-x}$, $a
> -1$) and Gaussian weight ($w(x)=e^{-x^2}$),  satisfy three-term recursion relation in the quaternion space.
From this, we derive generalized Christoffel-Darboux (GCD) formul\ae\
for kernel functions arising in the study of  the corresponding
orthogonal and symplectic ensembles of random $2N\times 2N$
matrices. Using the GCD formul\ae\, we calculate the level-densities
and prove that in the bulk of the spectrum, under appropriate
scaling, the eigenvalue correlations are universal. We also provide
evidence to show that there exists a mapping between skew-orthogonal
functions arising in the study of orthogonal and symplectic
ensembles of random matrices.
\end{abstract}
\maketitle

\section{Introduction}
\subsection{Random matrices}
Universality of eigenvalue correlations for different random matrix
ensembles and its application in real physical systems has
attracted both mathematicians and physicists in the last few decades
\cite{guhr,beenakker,ghosh,ghoshpandey,ghosh3,deift1,deift2,deift3,baik,bleher,deift4,deift5,deift6,deift7}.
From the mathematical point of view, the study of this universal behavior of
eigenvalue correlations for various ensembles of large random
matrices require (i) evaluation of certain kernel functions
involving orthogonal (unitary ensemble), skew-orthogonal (orthogonal
and symplectic ensemble) and bi-orthogonal (unitary two-matrix
ensemble) polynomials (ii) asymptotic analysis of these polynomials.
The rich literature available on orthogonal
\cite{szego,deift1,deift2,deift3,baik,bleher,deift4,deift5,deift6,deift7,eynard1,eynard11,plancherel}
and bi-orthogonal polynomials \cite{eynard2,eynard3,eynard4,Kap,KM}
corresponding to different weights have contributed a lot in our
understanding of unitary ensembles.

For orthogonal and symplectic ensembles, the problem of evaluating the kernel functions is
removed by the Tracy-Widom formalism, where the kernel functions are
expressed in terms of that of the unitary ensemble, involving
orthogonal polynomials \cite{tracy1,tracy2, tracy3, widom,sener}.
Results for orthogonal polynomials (OP) are then
used.

The second and perhaps a more enriching method is to evaluate the
kernel functions directly in terms of skew-orthogonal polynomials/skew-orthogonal functions
(SOP) and use asymptotic properties of these functions
\cite{dyson1}. For this, we need to develop the theory of SOP
\cite{ghosh,nagao1,nagao2,ghoshpandey,ghosh3,ghosh4,ghosh5,pandeyghosh,mehta,mehta1,eynard}
so that we can have further insight into orthogonal and symplectic
ensembles of random  matrices.

In this paper, we study statistical properties of orthogonal and
symplectic ensembles of random matrices with classical weight using
SOP. To do so, we evaluate the kernel functions (which we have
termed generalized Christoffel-Darboux sum or GCD) for the
corresponding orthogonal and symplectic ensembles of random
$2N\times 2N$ matrices. As $N\rightarrow\infty$, study of these
kernel functions require a knowledge of the asymptotic behavior of
the corresponding SOP. These can be derived by
solving the Riemann Hilbert problem \cite{ghosh3, pierce} for SOP.
The other (and perhaps easier) option is to obtain a finite term recursion relation between SOP
and OP and use known properties of the latter
(also see \cite{ghosh,ghoshpandey}) to obtain asymptotic results
for SOP. We use the second option.

OP satisfy three-term recursion relation,
irrespective of the weight function $w(x)$ w.r.t. which they are
defined. However recursion-relations satisfied by SOP depend on the corresponding weight.
This is because these functions, defined in the range
$[x_{1},x_{2}]$,  show skew-orthonormality w.r.t. their derivatives
(\ref{ortho1}). This brings into picture terms like $w'(x)/w(x)$
which gives rise to certain local behavior of the
recursion-relations. For example, for SOP
defined in the finite range $[x_{1},x_{2}]$, $w'(x)/w(x)$ may have
poles at the end points, which have to be dealt with in the recursion
relations. For SOP defined in the infinite range $[-\infty,\infty]$,
this can increase the number of terms in the recursion relations.

An obvious consequence of the local behavior of recursion relations
is that the GCD formul\ae\
 also depend on $w(x)$. In this paper, we
calculate the GCD sum corresponding to Jacobi weight, including the
limiting cases of associated Laguerre and Gaussian weight. GCD
formula for weight functions with  polynomial potential has
been derived in \cite{ghosh3} and is mentioned in this paper for
completeness. In \cite{ghosh,ghoshpandey} and \cite{mehta}, the
authors have found compact expressions for SOP in terms of
OP. In this paper, we use them and make further
developments of the asymptotic analysis of these SOP. Finally, using
these asymptotic results in the GCD formul\ae, we study the
statistical properties of the corresponding random matrix ensembles
in the bulk of the spectrum.

We also observe certain duality property between the two families of
SOP arising in the study of orthogonal and symplectic ensembles. (In fact, this
justifies further the use of SOP to ordinary OP
in studying these ensembles
\cite{stoj1,stoj1a,stoj2,deift1,deift2,nagao,tracy1,tracy2,tracy3,adler,widom}.)
However, this can at best be termed in physics literature as
``experimental observations''. We do not have a clear theoretical
understanding of this duality. The answer perhaps lies in the
existence of certain ortho-symplectic group which shares such
duality property. But the author's knowledge in this field is
severely limited.

We consider ensembles of $2N$ dimensional  matrices $H$ with
probability distribution
\begin{equation}
P_{\beta,N}(H)dH=\frac{1}{{\cal Z}_{\beta, N}}\exp[-[2{\rm
Tr}u(H)]]dH,
\end{equation}
where the matrix function $u(H)$ is defined by the power expansion
of the function $u(z)$. The parameter $\beta=1$, $2$ and $4$
correspond to ensembles invariant under orthogonal, unitary and
symplectic transformations. In this paper, we study $\beta=1$ and
$4$ cases.  The partition function is given by
\begin{equation}
\label{partition}
{\cal Z}_{\beta, N}:= \int_{H\in M^{\bt}_{2N}}\exp[-[2{\rm
Tr}u(H)]]dH=(2N)!\prod_{j=0}^{2N-1}g_{j}^{\bt},
\end{equation}
where $M^{\bt}_{2N}$ is a set of all $2N\times 2N$ real symmetric
($\beta=1$) and quaternion real self dual ($\beta=4$) matrices. $dH$
is the standard Haar measure. $g_j^{\bt}$ are normalization
constants for  SOP \cite{dyson1} corresponding to $\beta=1$ and $4$.




\subsection{Skew-orthogonal polynomials: relevance in orthogonal and symplectic
ensembles}


From the invariance of these ensembles under orthogonal ($\beta=1$)
and symplectic ($\beta=4$) transformation, joint probability of
eigenvalues ($x_{1},x_{2},\ldots, x_{2N}$) is  given by
\cite{mehta}:
\begin{equation}
P_{\beta,N}(x_{1},x_{2},\ldots, x_{2N})=\frac{1}{{\cal Z}_{\beta
N}}{|\Delta_{2N}(x_{1},x_{2},\ldots,
x_{2N})|}^{\beta}\prod_{j=1}^{2N}w(x_{j}),
\end{equation}
where $\Delta_{2N}(x_{1},x_{2},\ldots,
x_{2N})=\prod_{j<k}(x_{j}-x_{k})$ is the Vandermonde determinant.
The weight function $w(x)=\exp[-u(x)]$ is a non-zero and
non-negative function on the interval $[a,b]$ and have finite
moments.

The $n$-point correlation of eigenvalues is given by
\begin{eqnarray} \label{rn}
R_{n}^{(\beta)}(x_{1},\ldots,x_{n})=\frac{2N!}{(2N-n)!}\int
dx_{n+1}\ldots \int dx_{2N}P_{\beta,N}(x_{1},x_{2}\ldots,
x_{2N}),\qquad n=1,2,\ldots.
\end{eqnarray}
To evaluate such integrals, the joint probability distribution
$P_{\beta,N}(x_{1},x_{2},\ldots, x_{2N})$ is written in terms of
quaternion determinants (i.e. determinant of a matrix, each of whose
element is a $2\times 2$ quaternion) satisfying certain properties
\cite{mehta1}. Finally, using Dyson-Mehta theorem (page 152 of
\cite{mehta1}), one can calculate (\ref{rn}). For example, the two-point function
$R_{2}^{(\beta)}(x,y)$ and the level density $R_{1}^{(\beta)}(x)$ is given by
\begin{eqnarray}
\label{density}
R_{2}^{(\beta)}(x,y) &=& \left(\begin{array}{cc}
S^{(\beta)}_{2N}(x,y) & D^{(\beta)}_{2N}(x,y)     \\
I^{(\beta)}_{2N}(x,y)-\delta_{1,\beta}\epsilon (y-x) & S^{(\beta)}_{2N}(y,x)         \\
\end{array}\right);\qquad
R_{1}^{(\beta)}(x):={\rho}^{(\beta)}(x)=S_{2N}^{(\beta)}(x,x), \qquad \epsilon(r) =
\frac{|r|}{2r}.
\end{eqnarray}
Here $\delta$ is the kronecker delta. In terms of SOP
 $\phi_{n}^{\bt}(x)$  and $\psi_{n}^{\bt}(x)$, to be
defined in (\ref{quasipolynomial}) and (\ref{psi}) respectively,
they are expressed as:

\begin{eqnarray}
\nonumber\label{s} S^{(\beta)}_{2N}(x,y) &:=& \sum_{j,k=0}^{2N-1}
Z_{j,k}\phi^{\bt}_{j}(x)\psi^{\bt}_{k}(y)=
{{\widehat{\Psi}}^{(\beta)}}(y)\prod_{2N} \Phi^{(\beta)}(x)\\
&=& -{{\widehat{\Phi}}^{(\beta)}}(x)\prod_{2N} \Psi^{(\beta)}(y),
\end{eqnarray}
\begin{eqnarray}
\label{d} D^{(\beta)}_{2N}(x,y) &:=& -\sum_{j,k=0}^{2N-1}
Z_{j,k}\phi^{\bt}_{j}(x)\phi^{\bt}_{k}(y)=
{{\widehat{\Phi}}^{(\beta)}}(x)\prod_{2N} \Phi^{(\beta)}(y),\\
\label{i} I^{(\beta)}_{2N}(x,y) &:=& \sum_{j,k=0}^{2N-1}
Z_{j,k}\psi^{\bt}_{j}(x)\psi^{\bt}_{k}(y)=
-{{\widehat{\Psi}}^{(\beta)}}(x)\prod_{2N} \Psi^{(\beta)}(y), \\
S^{(\beta)}_{2N}(y,x) &=& {S^{\dagger}}^{\bt}_{2N}(x,y)=
{{\widehat{\Phi}}^{(\beta)}}(x)\prod_{2N} \Psi^{(\beta)}(y).
\end{eqnarray}
where
\begin{eqnarray}
\label{quasipolynomial} \phi_{n}^{\bt}(x) =
\frac{1}{\sqrt{g^{\bt}_n}}\pi^{\bt}_{n}(x)w(x),\qquad
 \pi^{\bt}_{n}(x) =
\sum^{n}_{k=0}c^{(n,\beta)}_{k}x^k,\qquad \beta=1,4,
\end{eqnarray}
are normalized SOP of order $n$.

Here, $\prod_{2N}={\rm diag}(\underbrace{{
1},\ldots,{1}}_{2N},0,\ldots, 0)$ is a diagonal matrix and
$\Phi^{\bt}(x)$  and $\Psi^{\bt}(x)$ are semi-infinite vectors:
\begin{eqnarray}
\label{phiinfty}
\Phi^{\bt}(x)={({\Phi_0^{\bt}}^t(x)\ldots{\Phi_n^{\bt}}^t(x)\ldots)}^t,\hspace{1cm}
\widehat{\Phi}^{\bt}(x)=
-{\Phi^{\bt}}^{t}(x)Z,\\
\label{psiinfty}
\Psi^{\bt}(x)={({\Psi_0^{\bt}}^t(x)\ldots{\Psi_n^{\bt}}^t(x)\ldots)}^t,\hspace{1cm}
\widehat{\Psi}^{\bt}(x)= -{\Psi^{\bt}}^{t}(x)Z,
\end{eqnarray}
with each entry a $2\times 1$ matrix:
\begin{eqnarray}
\label{twobyone} \Phi^{(\beta)}_n(x) = \left(\begin{array}{c}
\phi^{(\beta)}_{2n}(x)      \\
\phi^{(\beta)}_{2n+1}(x)   \\
\end{array}\right),
\hspace{1cm} \tphi^{(\beta)}_n(x) = {\left(\begin{array}{c}
\phi^{(\beta)}_{2n+1}(x)      \\
-\phi^{(\beta)}_{2n}(x)   \\
\end{array}\right)}^{t}.
\end{eqnarray}
(similar for $\Psi^{\bt}_{n}(x)$ and $\tpsi^{\bt}_n(x)$). The
anti-symmetric block-diagonal matrix $Z$ is given by
\begin{eqnarray}
Z &=& \left(\begin{array}{cc}
0 & 1     \\
-1 & 0         \\
\end{array}\right)
\dotplus \ldots \dotplus
\end{eqnarray}
such that $Z=-Z^t$ and $Z^2=-1$.

For
\begin{equation}
\label{psi} \Psi^{(4)}_{n}(x) =
  \Phi '^{(4)}_{n}(x),\qquad
\Psi^{(1)}_{n}(x)=\int_{\mathbb R}\Phi^{(1)}_{n}(y)\epsilon(x-y)dy,
\qquad n\in\mathbb N,
\end{equation}
 these polynomials satisfy skew-orthonormal relations w.r.t. the weight
 function $w^{2}(x)$\footnote{To observe the dual property among the two families
 of polynomials $\pi_{n}^{\bt}(x)$, $\beta=1$, $4$,
 we skew-orthonormalize them w.r.t.
 $w^{2}(x)$ (as in \ref{ortho1}) to {\it set} both
the families of SOP s on an equal footing. However to study the
statistical properties of  symplectic ensembles only, this is not
needed.}:

\begin{equation}
\label{ortho1} ({\Phi}^{\bt}_n,\tpsi_m^{\bt}) \equiv\int_{\mathbb
R}{\Phi}^{\bt}_n(x){\tpsi}_m^{\bt}(x) dx
=\delta_{nm}\left(\begin{array}{cc}
1 & 0     \\
0 & 1         \\
\end{array}\right),
\qquad n,m\in\mathbb N.
\end{equation}

Finally, from (\ref{psi}) and (\ref{s}), (\ref{d}) and (\ref{i}) we
get

\begin{eqnarray}
D^{{(1)}}_{2N}(x,y)=-\frac{\partial S^{{(1)}}_{2N}(x,y)}{\partial
y}, \qquad S^{{(1)}}_{2N}(x,y)=\frac{\partial
I^{{(1)}}_{2N}(x,y)}{\partial y},\\
I^{{(4)}}_{2N}(x,y)=\frac{\partial S^{{(4)}}_{2N}(x,y)}{\partial
x},\qquad S^{{(4)}}_{2N}(x,y)=-\frac{\partial
D^{{(4)}}_{2N}(x,y)}{\partial x}.
\end{eqnarray}
Thus a knowledge of the kernel function $S^{(\beta)}_{2N}(x,y)$ is
enough to calculate the correlation function. In this paper, we will
study the finite $N$ and large $N$ behavior of
$S^{(\beta)}_{2N}(x,y)$.

\vspace{0.3cm}

{\bf Outline of the paper}:

\vspace{0.05cm}

- In section 2, we calculate GCD formul\ae\ for the kernel function
$S_{2N}^{\bt}(x,y)$ (\ref{s}), $\beta=1$ and $4$,  corresponding to different weight.

\vspace{0.1cm}

-In section 3, we discuss the idea of duality that exists between
the two families of SOP arising in the study of orthogonal and
symplectic ensembles of random matrices.

\vspace{0.1cm}

-In section 4, we give a brief summary of some of the relevant
properties of classical OP which will be useful
in our study of the corresponding SOP.

\vspace{0.1cm}

- In section 5, we use results of section 2  to (i) obtain the
level densities (\ref{density}) for Jacobi and associated Laguerre
orthogonal ensembles, (ii) prove that in the bulk of the spectrum,
the kernel functions  $S_{2N}^{\bt}(x,y)/S_{2N}^{\bt}(x,x)$ and
hence the unfolded correlation functions for the above ensembles are
stationary and universal.

 \vspace{0.1cm}

 -In section 6, we repeat the same calculations for Jacobi symplectic and associated Laguerre
symplectic ensembles.

\vspace{0.1cm}

- Conclusion. \footnote{For random matrix ensembles with polynomial
potential, the special case of Gaussian ensemble ($d=1$), which is
also one of the limiting case of Jacobi ensemble, has been worked
out explicitly in \cite{ghosh3} using GCD formula, and hence not
repeated in this paper.}


\section{The Generalised Christoffel Darboux sum}

\subsection{Recursion Relations}


For polynomials with weight function
\begin{eqnarray}
w(x)={(x_{2}-x)}^{a}{(x-x_{1})}^{b},\qquad x_{1}, x_{2}\in\mathbb
{R}
\end{eqnarray}
skew-orthogonal in the finite interval $[x_{1},x_{2}]$, and having
finite moments, evaluation of $\psi^{(4)}_{n}(x)$ and
$\phi^{(1)}_{n}(x)$ will involve terms like $w'(x)/w(x)$ which have
poles at $x_{1}$ and $x_{2}$. Hence to obtain recursion relations we
expand ${\left[(x-x_1)(x_2-x)\Phi^{\bt}(x)\right]}'$ and
${\left[x(x-x_1)(x_2-x)\Phi^{\bt}(x)\right]}'$ in terms of
SOP $\Phi^{\bt}(x)$ (\ref{phiinfty}) and
introduce semi-infinite matrices $P^{(\beta)}$ and $R^{(\beta)}$
such that for $\beta=4$,

\begin{eqnarray}
(x-x_1)(x_2-x){(\Phi^{(4)}(x))}'\equiv
f(x){(\Phi^{(4)}(x))}'=P^{(4)}\Phi^{(4)}(x),\\
x(x-x_1)(x_2-x){(\Phi^{(4)}(x))}'\equiv
xf(x){(\Phi^{(4)}(x))}'=R^{(4)}\Phi^{(4)}(x).
\end{eqnarray}
For $\beta=1$, we get
\begin{eqnarray}
\label{genp1} {(x-x_1)(x_2-x)\Phi^{(1)}(x)} \equiv
f(x)\Phi^{(1)}(x)=P^{(1)}\Psi^{(1)}(x),\\
\label{genr1}
 {x(x-x_1)(x_2-x)\Phi^{(1)}(x)}\equiv xf(x)\Phi^{(1)}(x)=R^{(1)}\Psi^{(1)}(x).
\end{eqnarray}
Equations (\ref{genp1}) and (\ref{genr1}) are obtained by
multiplying the above expansion by $\epsilon(y-x)$ and integrating
by parts.

In this context, the Jacobi weight function is defined in the interval $[-1,1]$ by
\begin{equation}
\label{jacobi} w_{a,b}(x)={(1-x)}^{a}{(1+x)}^{b},\qquad a>-1,b>-1,
\end{equation}
where restrictions on $a$ and $b$ ensure that they have finite
moments.

Associated Laguerre weight function is defined in the interval
$[0,\infty]$ by
\begin{equation}
\label{laguerre} w_{a}(x)={x}^{a}e^{-x},\qquad a>-1,
\end{equation}
where restriction on $a$  ensures that they have finite moments.

Gaussian weight function is defined in the interval
$[-\infty,\infty]$ by
\begin{equation}
\label{gaussian} w(x)=e^{-x^2}.
\end{equation}

From here on, we will concentrate on these classical weight
functions and show that the corresponding SOP satisfy three-term
recursion relations in the $2\times 2$ quaternion space.

\subsection {Recursion relations for SOP with classical weight}

For classical weight, we expand ${\left[(f(x)\Phi^{\bt}(x)\right]}'$
and ${\left[x(f(x)\Phi^{\bt}(x)\right]}'$ in terms of
$\Phi^{(\beta)}(x)$. They satisfy the following recursion relations:

\begin{eqnarray}
 \label{PR4J} f(x)\Psi^{(4)}(x) &=&
P^{(4)}\Phi^{(4)}(x),{\hspace{1cm}}
xf(x)\Psi^{(4)}(x)=R^{(4)}\Phi^{(4)}(x),\\
\label{PR1J} f(x)\Phi^{(1)}(x) &=& P^{(1)}\Psi^{(1)}(x),\qquad
xf(x)\Phi^{(1)}(x)=R^{(1)}\Psi^{(1)}(x).
\end{eqnarray}
where $P^{\bt}$ and $R^{\bt}$ are semi-infinite tridiagonal
quaternion matrices. The semi-infinite vectors $\Phi^{\bt}(x)$ and
$\Psi^{\bt}(x)$ are given in (\ref{phiinfty}) and (\ref{psiinfty})
respectively. Now, $w'(x)/w(x)$ and hence $\Psi^{(4)}(x)$ and $\Phi^{(1)}(x)$  has singularity at $x\pm 1$ for
Jacobi, and at $x=0$ for associated Laguerre. To remove them, we have

\begin{eqnarray}
\label{fx}
f(x) &=& (1-x^2),\qquad {\rm Jacobi}\\
     &=& x, \qquad {\rm Associated\ Laguerre}\\
     &=& 1, \qquad {\rm Gaussian \ (also\ true\ for\ any\ polynomial\ weight)}.
\end{eqnarray}
In other words, SOP satisfy three-term recursion relation in
the quaternion space and is given by:
\begin{eqnarray}
\label{P4Jquat} f(x)\Psi_{n}^{(4)}(x) &=&
{\bf P}_{n,n+1}^{(4)}\Phi_{n+1}^{(4)}(x)+{\bf P}_{n,n}^{(4)}\Phi_{n}^{(4)}(x)+{\bf P}_{n,n-1}^{(4)}\Phi_{n-1}^{(4)}(x),\\
\label{R4Jquat} xf(x)\Psi_{n}^{(4)}(x) &=& {\bf
R}_{n,n+1}^{(4)}\Phi_{n+1}^{(4)}(x)+{\bf
R}_{n,n}^{(4)}\Phi_{n}^{(4)}(x)+{\bf
R}_{n,n-1}^{(4)}\Phi_{n-1}^{(4)}(x).\\
 \label{P1Jquat}
f(x)\Phi_{n}^{(1)}(x) &=&
{\bf P}_{n,n+1}^{(1)}\Psi_{n+1}^{(1)}(x)+{\bf P}_{n,n}^{(1)}\Psi_{n}^{(1)}(x)+{\bf P}_{n,n-1}^{(1)}\Psi_{n-1}^{(1)}(x),\\
\label{R1Jquat} xf(x)\Phi_{n}^{(1)}(x) &=& {\bf
R}_{n,n+1}^{(1)}\Psi_{n+1}^{(1)}(x)+{\bf
R}_{n,n}^{(1)}\Psi_{n}^{(1)}(x)+{\bf
R}_{n,n-1}^{(1)}\Psi_{n-1}^{(1)}(x),
\end{eqnarray}
where $\Phi_{n}^{\bt}(x)$ and $\Psi_{n}^{\bt}(x)$ are given in
(\ref{twobyone}) and ${\bf P}_{j,k}^{\bt}$ and ${\bf R}_{j,k}^{\bt}$
are $2\times 2$ quaternions. Equation
(\ref{P4Jquat})-(\ref{R1Jquat}) can be proved directly using the
skew-orthogonal relation (\ref{ortho1}). We leave it as an exercise.
In this paper, we will give an alternative proof by showing that the
semi-infinite matrices $P^{\bt}$ and $R^{\bt}$ are tridiagonal (in
the quaternion sense) and anti-self dual.

In terms of the elements of the quaternion matrices, (\ref{P4Jquat})
and (\ref{R4Jquat}) can be written as:
\begin{eqnarray}
\nonumber f(x)\left(\begin{array}{cc}
\psi^{(4)}_{2n}(x)      \\
\psi^{(4)}_{2n+1}(x)     \\
\end{array}\right)
\label{p4Jquatelem} &=& \left(\begin{array}{cc}
0 & 0     \\
P^{(4)}_{2n+1,2n+2} & 0         \\
\end{array}\right)
\left(\begin{array}{cc}
\phi^{(4)}_{2n+2}(x)      \\
\phi^{(4)}_{2n+3}(x)          \\
\end{array}\right)
+ \left(\begin{array}{cc}
P^{(4)}_{2n,2n} & P^{(4)}_{2n,2n+1}     \\
P^{(4)}_{2n+1,2n} & P^{(4)}_{2n+1,2n+1}         \\
\end{array}\right)
\left(\begin{array}{cc}
\phi^{(4)}_{2n}(x)      \\
\phi^{(4)}_{2n+1}(x)          \\
\end{array}\right)\\
&& + \left(\begin{array}{cc}
0 & 0     \\
P^{(4)}_{2n+1,2n-2} & 0         \\
\end{array}\right)
\left(\begin{array}{cc}
\phi^{(4)}_{2n-2}(x)      \\
\phi^{(4)}_{2n-1}(x)          \\
\end{array}\right)
\end{eqnarray}

and
\begin{eqnarray}
\nonumber xf(x)\left(\begin{array}{cc}
\psi^{(4)}_{2n}(x)      \\
\psi^{(4)}_{2n+1}(x)     \\
\end{array}\right)
\label{r4Jquatelem} &=& \left(\begin{array}{cc}
R^{(4)}_{2n,2n+2} & 0     \\
R^{(4)}_{2n+1,2n+2} & R^{(4)}_{2n+1,2n+3}         \\
\end{array}\right)
\left(\begin{array}{cc}
\phi^{(4)}_{2n+2}(x)      \\
\phi^{(4)}_{2n+3}(x)          \\
\end{array}\right)
+ \left(\begin{array}{cc}
R^{(4)}_{2n,2n} & R^{(4)}_{2n,2n+1}     \\
R^{(4)}_{2n+1,2n} & R^{(4)}_{2n+1,2n+1}         \\
\end{array}\right)
\left(\begin{array}{cc}
\phi^{(4)}_{2n}(x)      \\
\phi^{(4)}_{2n+1}(x)          \\
\end{array}\right)\\
&& + \left(\begin{array}{cc}
R^{(4)}_{2n,2n-2} & 0     \\
R^{(4)}_{2n+1,2n-2} & R^{(4)}_{2n+1,2n-1}         \\
\end{array}\right)
\left(\begin{array}{cc}
\phi^{(4)}_{2n-2}(x)      \\
\phi^{(4)}_{2n-1}(x)          \\
\end{array}\right).
\end{eqnarray}

For $\beta=1$, we get similar relations, where $\Phi^{(4)}$ and
$\Psi^{(4)}$ are replaced by $\Psi^{(1)}$ and $\Phi^{(1)}$
respectively.

For the polynomial weight, the semi-infinite matrices $P^{\bt}$ and
$R^{\bt}$ have $d$ quaternion bands above and below the diagonal
\cite{ghosh3}. Thus the Gaussian ($d=1$) SOP,
like the Jacobi and associated Laguerre functions, satisfy three
term recursion in the quaternion space.

{\bf Note:} Here, we would like to mention that unlike OP,
the Jacobi matrix $Q^{\bt}$ coming from the relation
$x\Phi^{\bt}(x)=Q^{\bt}\Phi^{\bt}(x)$, $\beta=1,4$, holds little
importance as they do not have finite bands below the diagonal.

\subsection{Proof}
To prove that the SOP corresponding to
classical weight satisfy three-term recursion in the quaternion
space, we will prove that the matrices $P^{(4)}$ and $R^{(4)}$ for
Jacobi weight are anti-self dual. We use the scalar products
\begin{eqnarray}
\sum_{j} P^{(4)}_{n,j}Z_{j,n} &=&
\left((1-x^2)\psi^{(4)}_{n}(x),\psi^{(4)}_{m}(x)\right)=\sum_{j}
P^{(4)}_{m,j}Z_{j,n},\\
\sum_{j} R^{(4)}_{n,j}Z_{j,n} &=&
\left(x(1-x^2)\psi^{(4)}_{n}(x),\psi^{(4)}_{m}(x)\right)=\sum_{j}
R^{(4)}_{m,j}Z_{j,n}.
\end{eqnarray}
Similarly, using $\left(x\psi^{(4)}_{n}(x),\psi^{(4)}_{m}(x)\right)$
and $\left(x^2\psi^{(4)}_{n}(x),\psi^{(4)}_{m}(x)\right)$ for
associated Laguerre weight and
$\left(\psi^{(4)}_{n}(x),\psi^{(4)}_{m}(x)\right)$ and
$\left(x\psi^{(4)}_{n}(x),\psi^{(4)}_{m}(x)\right)$ for polynomial
weight for $\beta=4$ and replacing $\psi^{(4)}(x)$ by
$\phi^{(1)}(x)$ for $\beta=1$, we get

\begin{eqnarray}
\label{antidual} P^{(\beta)}=-{P^{(\beta)}}^{D}, \qquad
R^{(\beta)}=-{R^{(\beta)}}^{D},
\end{eqnarray}
where dual of a matrix is defined as
\begin{eqnarray}
A^D:=-ZA^{t}Z.
\end{eqnarray}

It is straightforward to see that $P^{\bt}$ and $R^{\bt}$ have
finite bands (one in the case of SOP defined w.r.t. the classical
weight functions) above the diagonal. Equation (\ref{antidual})
ensures that they also have the same number of bands (where each
entry is a $2\times 2$ quaternion \cite{mehta1}) below the diagonal.
This completes the proof.

\subsection{Generalized Christoffel Darboux sum}

In this subsection, we generalize the results given in \cite{ghosh3}
to include  GCD sum for both classical weights as well as weight
functions with polynomial potential.

With $f(y)$ given in (\ref{fx}), we use (\ref{s}) and (\ref{PR4J})
to get,

\begin{eqnarray}
\nonumber f(y)S^{(4)}_{2N}(x,y)-f(x)S^{(4)}_{2N}(y,x) &=&
f(y)\left[{{\Phi}^{(4)}}^{t}(x)\prod_{2N}Z\prod_{2N}\Psi^{(4)}(y)\right]
+f(x)\left[{{\Psi}^{(4)}}^{t}(x)\prod_{2N}Z\prod_{2N}\Phi^{(4)}(y) \right]\\
\nonumber &=&
\left[-{{\Phi}^{(4)}}^{t}(x)ZZ\prod_{2N}Z\prod_{2N}P^{(4)}\Phi^{(4)}(y)\right]
+\left[{{\Phi}^{(4)}}^{t}(x)ZZ{P^{(4)}}^{t}ZZ\prod_{2N}Z\prod_{2N}\Phi^{(4)}(y) \right]\\
\nonumber &=&
\left[-{{\widehat{\Phi}}}^{(4)}(x)\prod_{2N}P^{(4)}\Phi^{(4)}(y)\right]
+\left[{{\widehat{\Phi}}}^{(4)}(x)P^{(4)}\prod_{2N}\Phi^{(4)}(y) \right]\\
&=&
{{\widehat{\Phi}}}^{(4)}(x)\left[P^{(4)},\prod_{2N}\right]\Phi^{(4)}(y).
\end{eqnarray}

Similarly,

\begin{eqnarray}
\nonumber yf(y)S^{(4)}_{2N}(x,y)-xf(x)S^{(4)}_{2N}(y,x) &=&
yf(y)\left[{{\Phi}^{(4)}}^{t}(x)\prod_{2N}Z\prod_{2N}\Psi^{(4)}(y)\right]
+xf(x)\left[{{\Psi}^{(4)}}^{t}(x)\prod_{2N}Z\prod_{2N}\Phi^{(4)}(y) \right]\\
\nonumber &=&
\left[-{{\Phi}^{(4)}}^{t}(x)ZZ\prod_{2N}Z\prod_{2N}R^{(4)}\Phi^{(4)}(y)\right]
+\left[{{\Phi}^{(4)}}^{t}(x)ZZ{R^{(4)}}^{t}ZZ\prod_{2N}Z\prod_{2N}\Phi^{(4)}(y) \right]\\
\nonumber &=&
\left[-{{\widehat{\Phi}}}^{(4)}(x)\prod_{2N}R^{(4)}\Phi^{(4)}(y)\right]
+\left[{{\widehat{\Phi}}}^{(4)}(x)R^{(4)}\prod_{2N}\Phi^{(4)}(y) \right]\\
&=&
{{\widehat{\Phi}}}^{(4)}(x)\left[R^{(4)},\prod_{2N}\right]\Phi^{(4)}(y).
\end{eqnarray}
Combining the two, GCD formula for symplectic ensembles of random
matrices with classical weight is given by

\begin{eqnarray}
\label{gcdjac4} S^{(4)}_{2N}(x,y) &=&
\frac{{{\widehat{\Phi}}}^{(4)}(x)[{\overline
R}^{(4)}(x),\prod_{2N}]{\Phi^{(4)}(y)}}{f(y)(y-x)},\qquad N\geq 1.
\end{eqnarray}

For the corresponding orthogonal ensembles ($\beta=1$), GCD formula
is derived using similar technique. From (\ref{s}) and (\ref{PR1J}),
we get
\begin{eqnarray}
\nonumber f(x)S^{(1)}_{2N}(x,y)-f(y)S^{(1)}_{2N}(y,x) &=&
f(x)\left[{{\Phi}^{(1)}}^{t}(x)\prod_{2N}Z\prod_{2N}\Psi^{(1)}(y)\right]
+f(y)\left[{{\Psi}^{(1)}}^{t}(x)\prod_{2N}Z\prod_{2N}\Phi^{(1)}(y)\right]\\
&=& {\widehat{\Psi}}^{(1)}(x)
\left[P^{(1)},\prod_{2N}\right]{\Psi^{(1)}(y)},
\end{eqnarray}
and
\begin{eqnarray}
\nonumber xf(x)S^{(1)}_{2N}(x,y)-yf(y)S^{(1)}_{2N}(y,x) &=&
xf(x)\left[{{\Phi}^{(1)}}^{t}(x)\prod_{2N}Z\prod_{2N}\Psi^{(1)}(y)\right]
+yf(y)\left[{{\Psi}^{(1)}}^{t}(x)\prod_{2N}Z\prod_{2N}\Phi^{(1)}(y)\right],\\
&=& {\widehat{\Psi}}^{(1)}(x)
\left[R^{(1)},\prod_{2N}\right]{\Psi^{(1)}(y)}.
\end{eqnarray}
Combining the two, the GCD formula for classical orthogonal
ensembles is given by
\begin{eqnarray}
\label{gcdjac1} S^{(1)}_{2N}(x,y) = \frac{{\widehat{\Psi}}^{(1)}(x)
[{\overline
R}^{(1)}(y),\prod_{2N}]{\Psi^{(1)}(y)}}{f(x)(x-y)},\qquad N\geq 1.
\end{eqnarray}

Here
\begin{equation}
{\overline R}^{\bt}(x)=R^{\bt}-xP^{\bt},\qquad \beta=1,4,
\end{equation}
is different for different weights.

For example, GCD matrix for the Jacobi symplectic ensemble
(including the associated Laguerre and Gaussian symplectic ensemble)
has the following structure:

\begin{eqnarray}
\nonumber && {\widehat{\Phi}}^{(4)}(x) \left[{\overline
R}^{(4)}(x),\prod_{2N}\right]{\Phi^{(4)}(y)}\\
= &&\nonumber {\left(\begin{array}{cccccc}
\phi^{(4)}_{1}(x) \\
-\phi^{(4)}_{0}(x) \\
\vdots \\
\end{array}\right)}^{t}
\left(\begin{array}{cccccc}
0&0&0&0&0&0 \\
0&0&0&\vdots&\vdots&0 \\
0&0&0&-{R}_{2N-2,2N}^{(4)}&0&0 \\
0&0&0&-{\overline R}_{2N-1,2N}^{(4)}(x)
                     &-{R}_{2N-1,2N+1}^{(4)}&0 \\
0& -{R}_{2N-1,2N+1}^{(4)} &0&0&0&0\\
0&{\overline R}_{2N-1,2N}^{(4)}(x)&-{R}_{2N-2,2N}^{(4)} & 0 &\ldots & 0 \\
0&0&0&0&0&0 \\
0&0&0&\vdots&\vdots&0 \\
\end{array}\right)
\left(\begin{array}{cccccc}
\phi^{(4)}_{0}(y) \\
\phi^{(4)}_{1}(y) \\
\vdots \\
\end{array}\right),\\
&=& \nonumber
R^{(4)}_{2N-2,2N}\left[\phi^{(4)}_{2N}(x)\phi^{(4)}_{2N-1}(y)
-\phi^{(4)}_{2N}(y)\phi^{(4)}_{2N-1}(x)\right]+
R^{(4)}_{2N-1,2N+1}\left[\phi^{(4)}_{2N-2}(x)\phi^{(4)}_{2N+1}(y)
-\phi^{(4)}_{2N-2}(y)\phi^{(4)}_{2N+1}(x)\right]\\
\label{rgcd4}  && +
\left(R^{(4)}_{2N-1,2N}-xP^{(4)}_{2N-1,2N}\right)\left[\phi^{(4)}_{2N-2}(x)\phi^{(4)}_{2N}(y)
-\phi^{(4)}_{2N-2}(y)\phi^{(4)}_{2N}(x)\right].
\end{eqnarray}

Similarly, for classical orthogonal ensemble, the GCD matrix has the
following structure:
\begin{eqnarray}
\nonumber  && {\widehat{\Psi}}^{(1)}(x) \left[{\overline
R}^{(1)}(y),\prod_{2N}\right]{\Psi^{(1)}(y)} \\
&=& \nonumber
R^{(1)}_{2N-2,2N}\left[\psi^{(1)}_{2N}(x)\psi^{(1)}_{2N-1}(y)
-\psi^{(1)}_{2N}(y)\psi^{(1)}_{2N-1}(x)\right]+
R^{(1)}_{2N-1,2N+1}\left[\psi^{(1)}_{2N-2}(x)\psi^{(1)}_{2N+1}(y)
-\psi^{(1)}_{2N-2}(y)\psi^{(1)}_{2N+1}(x)\right]\\
\label{rgcd1}  && +
\left(R^{(1)}_{2N-1,2N}-yP^{(1)}_{2N-1,2N}\right)\left[\psi^{(1)}_{2N-2}(x)\psi^{(1)}_{2N}(y)
-\psi^{(1)}_{2N-2}(y)\psi^{(1)}_{2N}(x)\right].
\end{eqnarray}

\section{Duality}

Duality between the two families of SOP
arising in the study of orthogonal ($\beta=1$) and symplectic
($\beta=4$) ensembles of random matrices, was predicted in
\cite{ghosh3} and \cite{ghosh4}. In this section, we show the
existence of such duality between the two families of SOP corresponding to classical weight, i.e.
$\Phi^{(4)}(x)\mapsto \Psi^{(1)}(x)$ and $\Psi^{(4)}(x)\mapsto
\Phi^{(1)}(x)$. For this, we derive recursion relations between the two families of SOP with their corresponding OP.
 Apart from demonstrating duality, this technique simplifies the derivation
of asymptotic results of the SOP.

We expand functions $\Phi^{(4)}_{m}(x)$ and $\Psi^{(1)}_{m}(x)$,
$m\geq 1$,  skew-orthogonal in the range  $[x_{1},x_{2}]$, in a
suitable basis of OP such that:

(i) their derivatives are continuous in the range $[x_{1},x_{2}]$
and vanish at the end-points.

(ii) $\phi^{(4)}_{m}(x)$ and
${(\psi^{(1)}_{m}(x))}'\equiv\phi^{(1)}_{m}(x)$ can be written as
$w(x){\pi}^{\bt}_{m}(x)$.

(iii) $\Phi^{(4)}_{m}(x)$ and $\Psi^{(1)}_{m}(x)$ are
skew-orthonormal in the range  $[x_{1},x_{2}]$ w.r.t. their
derivatives.

We expand Jacobi SOP $\phi^{(4)}_{m}(x)$ and
$\psi^{(1)}_{m}(x)$ in terms of Jacobi OP
$P_{j}^{2a+1,2b+1}(x)$, orthogonal w.r.t. the weight function
$w_{2a+1,2b+1}(x)$ (see \ref{jacobi}). For associated Laguerre SOP,
we expand in terms of $L_{j}^{2a+1}(x)$,
orthogonal w.r.t. the weight function $w_{2a+1}(x)$ (see
\ref{laguerre}) while for Gaussian SOP, the
basis chosen is $H_{j}(x)$ orthogonal w.r.t. the weight $e^{-x^2}$.
The choice of such basis ensures that conditions (i-iii) are
satisfied.

\subsection{Jacobi SOP}

Jacobi SOP corresponding to orthogonal and
symplectic ensembles are given below. We see that there exists a
relation between $\psi^{(1)}_{m}(x)$ and $\phi^{(4)}_{m}(x)$ and
their derivatives. We will use
the following identities:
\begin{eqnarray}
\label{wab1} (1-x^2)w_{a,b}(x)=w_{a+1,b+1}(x),\qquad
w_{a,b}(x)w_{a+1,b+1}(x)=w_{2a+1,2b+1}(x),
\end{eqnarray}

For $\beta=1$, with $\Phi_{m}^{(1)}(x)$ and $\Psi_{m}^{(1)}(x)$
satisfying conditions (i-iii), we have for $m\geq 1$,
\begin{eqnarray}
\label{psijac1}
&& {(g^{(1)}_{2m})}^{1/2}\psi^{(1)}_{2m+1}(x)= w_{a+1,b+1}(x)P^{2a+1,2b+1}_{2m}(x),\\
\label{phijac1odd}&& {(g^{(1)}_{2m})}^{1/2}\phi^{(1)}_{2m+1}(x)=
w_{a,b}(x)[A_{2m+1}P^{2a+1,2b+1}_{2m+1}(x)-B_{2m-1}P^{2a+1,2b+1}_{2m-1}(x)],\\
\label{psijac1even} && {(g^{(1)}_{2m})}^{1/2}\psi^{(1)}_{2m}(x)=
\frac{w_{a+1,b+1}(x)}{A_{2m}}P^{2a+1,2b+1}_{2m-1}(x)
+\gamma^{(2m)}_{2m-2}\psi^{(1)}_{2m-2}(x), \\
\label{phijac1} && {(g^{(1)}_{2m})}^{1/2}\phi^{(1)}_{2m}(x)=
w_{a,b}(x)P^{2a+1,2b+1}_{2m}(x),
\end{eqnarray}
with ${(g^{(1)}_{0})}^{1/2}\psi^{(1)}_{0}(x)=
\int\epsilon(x-y)w_{a,b}(y)dy$ and
\begin{eqnarray}
g^{(1)}_{2m}=g^{(1)}_{2m+1}=h^{2a+1,2b+1}_{2m},\qquad m=0,1,\ldots.
\end{eqnarray}

For $\beta=4$, with $\Phi_{m}^{(4)}(x)$ and $\Psi_{m}^{(4)}(x)$
satisfying conditions (i-iii), we have for $m\geq 1$,
\begin{eqnarray}
\label{dual41}
&& {(g^{(4)}_{2m})}^{1/2}\phi^{(4)}_{2m+1}(x) = w_{a+1,b+1}(x)P^{2a+1,2b+1}_{2m-1}(x),\\
&& {(g^{(4)}_{2m})}^{1/2}\psi^{(4)}_{2m+1}(x)=
w_{a,b}(x)[A_{2m}P^{2a+1,2b+1}_{2m}(x)-B_{2m-2}P^{2a+1,2b+1}_{2m-2}(x)],\\
&& {(g^{(4)}_{2m})}^{1/2}\phi^{(4)}_{2m}(x)
=-\frac{w_{a+1,b+1}(x)}{A_{2m-1}}P^{2a+1,2b+1}_{2m-2}(x)
+\gamma^{(2m-1)}_{2m-3}\phi^{(4)}_{2m-2}(x),\\
\label{duality42} && {(g^{(4)}_{2m})}^{1/2}\psi^{(4)}_{2m}(x)=
-w_{a,b}(x)P^{2a+1,2b+1}_{2m-1}(x),
\end{eqnarray}
with
\begin{eqnarray}
&& g^{(4)}_{2m}=g^{(4)}_{2m+1}=h^{2a+1,2b+1}_{2m-1},\qquad
m=1,2,\ldots.
\end{eqnarray}
where
\begin{eqnarray}
\label{gamma}
\gamma^{(j)}_{j-2}\equiv\gamma_{j}=\frac{(j+2a+2)(j+2b+2)}{(j+2)(j+2a+2b+4)},\qquad
   A_{j}  =-\frac{j(j+2a+2b+2)}{(2j+2a+2b+1)},\qquad
     B_{j} = -\frac{(j+2a+2)(j+2b+2)}{(2j+2a+2b+5)},
\end{eqnarray}
\begin{eqnarray}
\label{abh}
     \frac{B_{j-1}}{A_{j}}h_{j-1}^{2a+1,2b+1} = h_{j}^{2a+1,2b+1},\qquad B_{-l}=0,\qquad l=1,2,\ldots.
\end{eqnarray}
$\phi^{(4)}_{0}(x)$ and $\phi^{(4)}_{1}(x)$ can be calculated using
(\ref{def1}) and Gram-Schmidt method for SOP. Here we note that
SOP for $\beta=4$ is lower than that of
$\beta=1$ by an order $1$.

\subsection{Associated Laguerre SOP}

Associated Laguerre ensembles of random matrices can and does play a
significant role in describing real physical systems \cite{sener}. There exists a simple duality relation between
the SOP $\Psi_{m}^{(1)}(x)$ and
$\Phi_{m}^{(4)}(x)$, their derivatives and the normalization constant for $m\geq 1$:

\begin{eqnarray}
\label{dualityrelation}
\Psi_{m}^{(1)}(x)=-\sigma_{3}\Phi_{m}^{(4)}(x);\qquad
\Phi_{m}^{(1)}(x)=-\sigma_{3}\Psi_{m}^{(4)}(x);\qquad g_{m}^{(4)}=g_{m}^{(1)},\qquad{\rm where}\qquad
\sigma_{3} &=& \left(\begin{array}{cc}
1 & 0     \\
0 & -1         \\
\end{array}\right)
\end{eqnarray}
is the Pauli matrix.\footnote {Using (\ref{dualityrelation}) in
(\ref{partition}) we can see that the partition functions corresponding to Laguerre weight  also share a duality relation.}

We now present the results for the SOP
corresponding to associated Laguerre weight. We use
\begin{eqnarray}
2^{2a+1}w_{a}(x)w_{a+1} (x) = w_{2a+1}(2x),\qquad
         w_{a+1}(x) = xw_{a}(x),
\end{eqnarray}
the latter vanishing at $x=0$ for all $a>-1$.

For $\beta=1$, with
${(g^{(1)}_{m})}^{1/2}\Phi_{m}^{(1)}(x)=w_{a}(x)\pi_{m}^{(1)}(x)$,
we have for $m\geq 1$,
\begin{eqnarray}
\label{psi1lagodd}
&& {(g^{(1)}_{2m+1})}^{1/2}\psi^{(1)}_{2m+1}(x)= 2^{a+3/2}w_{a+1}(x)L^{2a+1}_{2m}(2x)\\
\label{phi1lagodd} && {(g^{(1)}_{2m+1})}^{1/2}\phi^{(1)}_{2m+1}(x)=
2^{a+1/2}w_{a}(x)[A^{L}_{2m+1}L^{2a+1}_{2m+1}(2x)-B^{L}_{2m-1}L^{2a+1}_{2m-1}(2x)],\qquad B_{-1}^{L}=0,\\
\label{psi1lageven} && {(g^{(1)}_{2m})}^{1/2}\psi^{(1)}_{2m}(x)=
2^{a+3/2}\frac{w_{a+1}(x)}{A^{L}_{2m}}L^{2a+1}_{2m-1}(x)
+\gamma^{(2m)}_{2m-2}\psi^{(1)}_{2m-2}(x),\qquad m\neq 0,\\
\label{phi1lageven}
&& {(g^{(1)}_{2m})}^{1/2}\phi^{(1)}_{2m}(x)= 2^{a+1/2}w_{a}(x)L^{2a+1}_{2m}(2x),\\
&& g^{(1)}_{2m}=g^{(1)}_{2m+1}=h_{2m}^{2a+1},\qquad m=0,1,\ldots.
\end{eqnarray}
where, ${(g^{(1)}_{0})}^{1/2}\psi^{(1)}_{0}(x)=
2^{a+1/2}\int_{0}^{\infty}\epsilon(x-y)w_{a}(y)dy $.

 For $\beta=4$, with
${(g^{(4)}_{m})}^{1/2}\Phi_{m}^{(4)}(x)=w_{a}(x)\pi_{m}^{(4)}(x)$,
we have
\begin{eqnarray}
\label{dual43}
&& {(g^{(4)}_{2m+1})}^{1/2}\phi^{(4)}_{2m+1}(x) = 2^{a+3/2}w_{a+1}(x)L^{2a+1}_{2m}(2x)\\
&& {(g^{(4)}_{2m+1})}^{1/2}\psi^{(4)}_{2m+1}(x)=
2^{a+1/2}w_{a}(x)[A^{L}_{2m+1}L^{2a+1}_{2m+1}(2x)-B^{L}_{2m-1}L^{2a+1}_{2m-1}(2x)]\\
&& {(g^{(4)}_{2m})}^{1/2}\phi^{(4)}_{2m}(x)
=-2^{a+3/2}\frac{w_{a+1}(x)}{A^{L}_{2m}}L^{2a+1}_{2m-1}(2x)
+\gamma^{(2m-1)}_{2m-3}\phi^{(4)}_{2m-2}(x),\qquad m\neq 0,\\
\label{dual44} && {(g^{(4)}_{2m})}^{1/2}\psi^{(4)}_{2m}(x)=
-2^{a+1/2}w_{a}(x)L^{2a+1}_{2m}(2x)\\
&& \label{duality45}
g^{(4)}_{2m}=g^{(4)}_{2m+1}=h_{2m}^{2a+1},\qquad m=0,1,\ldots.
\end{eqnarray}
where ${(g^{(4)}_{0})}^{1/2}\phi^{(4)}_{0}(x)\propto w_{a}(x)$.

Here $\gamma^{(j)}_{j-2}\equiv \gamma_{j}^{L}$, $A_{j}^{L}$ and
$B_{j}^{L}$ are given by

\begin{eqnarray}
\label{gammal} \gamma_{j}^{L}&=&\frac{(j+2a+2)}{(j+2)},\qquad
   A_{j}^{L}=j,\qquad
     B_{j}^{L} = j+2a+2,\\
\label{abhL} \frac{B^{L}_{j-1}}{A^{L}_{j}}h^{2a+1}_{j-1} &=&
h^{2a+1}_{j},\qquad j\geq 1.
\end{eqnarray}
Here, the identity
\begin{equation}
\label{derivativel} \frac{d}{dx}\{w_{a+1}(x)L_{j}^{(2a+1)}(2x)\}
                       =\frac{1}{2}w_{a}(x)\{A_{j+1}^{L}L_{j+1}^{(2a+1)}(2x)
                               -B_{j-1}^{L}L_{j-1}^{(2a+1)}(2x)\},
\end{equation}
is used to derive (\ref{psi1lagodd}) and (\ref{psi1lageven}) from
(\ref{phi1lagodd}) and (\ref{phi1lageven}) respectively.

\subsection{Gaussian SOP}

For Gaussian weight, with $w(x)=e^{-x^2/2}$ and $\beta=1$, there
exists a duality relation between the SOP
$\Psi_{m}^{(1)}(x)$ and $\Phi_{m}^{(4)}(x)$ and their derivatives
for $m\geq 1$, and is given by

\begin{eqnarray}
&& {(g^{(1)}_{2m+1})}^{1/2}\psi^{(1)}_{2m+1}(x)= w(x)H_{2m}(x)\\
&& {(g^{(1)}_{2m+1})}^{1/2}\phi^{(1)}_{2m+1}(x)=
w(x)[-(1/2)H_{2m+1}(x)+2m H_{2m-1}(x)]\\
&& {(g^{(1)}_{2m})}^{1/2}\psi^{(1)}_{2m}(x)= -2w(x)H_{2m-1}(x)
+2(2m-1)\psi^{(1)}_{2m-2}(x),\qquad m\neq 0.\\
&& {(g^{(1)}_{2m})}^{1/2}\phi^{(1)}_{2m}(x)= w(x)H_{2m}(x)\\
&& g^{(1)}_{2m}=g^{(1)}_{2m+1}=h_{2m},\qquad m=0,1,\ldots.
\end{eqnarray}
with ${(g^{(1)}_{0})}^{1/2}\psi^{(1)}_{0}(x)=
\int_{-\infty}^{\infty}\epsilon(x-y)w(y)dy $ and $H_{j}(x)$ the
ordinary Hermite polynomials.

For $\beta=4$, we have
\begin{eqnarray}
\label{dual45}
&& {(g^{(4)}_{2m+1})}^{1/2}\phi^{(4)}_{2m+1}(x) = w(x)H_{2m+1}(x)\\
&& {(g^{(4)}_{2m+1})}^{1/2}\psi^{(4)}_{2m+1}(x)=
w(x)[-(1/2)H_{2m+2}(x)+(2m+1) H_{2m}(x)]\\
&& {(g^{(4)}_{2m})}^{1/2}\phi^{(4)}_{2m}(x) =2w(x)H_{2m}(x)
+4m \phi^{(4)}_{2m-2}(x)\\
\label{dual46} && {(g^{(4)}_{2m})}^{1/2}\psi^{(4)}_{2m}(x)= -w(x)H_{2m+1}(x)\\
&& g^{(4)}_{2m}=g^{(4)}_{2m+1}=h_{2m+1},\qquad m=0,1,\ldots.
\end{eqnarray}
Thus for the Gaussian weight,  SOP for
$\beta=4$ is higher than that of $\beta=1$ by an order $1$. This is
exactly the opposite of what we saw for Jacobi SOP.

\subsection{Proof}

The Jacobi SOP $\phi_{j}^{(4)}(x)$ and
$\psi_{j}^{(1)}(x)$ can be written as

\begin{eqnarray}
{(g^{(4)}_{j})}^{1/2}\phi^{(4)}_{j}(x) &=&
\gamma_{j-1}^{(j-1)}w_{a+1,b+1}P_{j-2}^{2a+1,2b+1}(x)+\sum_{k=0}^{j-1}
\gamma_{k-1}^{(j-1)}{(g^{(4)}_{k})}^{1/2}\phi^{(4)}_{k}(x), \qquad
j\geq 2,\qquad
\gamma_{j-1}^{(j-1)}\neq 0,\\
{(g^{(1)}_{j})}^{1/2}\psi^{(1)}_{j}(x) &=&
\gamma_{j}^{(j)}w_{a+1,b+1}P_{j-1}^{2a+1,2b+1}(x)+\sum_{k=0}^{j-1}
\gamma_{k}^{(j)}{(g^{(1)}_{k})}^{1/2}\psi^{(1)}_{k}(x),\qquad j\geq
1,\qquad \gamma_{j}^{(j)}\neq 0,
\end{eqnarray}
such that they satisfy conditions (i-iii). For example,
$w_{a+1,b+1}(x)$ is used instead of $w_{a,b}(x)$ in order to satisfy
condition (i). The order of the OP are fixed by
(ii), while (iii) fixes $\gamma^{(j)}_{k}$.

Differentiating and using the identity
\begin{equation}
\label{derivative}
\frac{d}{dx}\{w_{a+1,b+1}(x)P_{j}^{2a+1,2b+1}(x)\}=
                                   w_{a,b}(x)\{A_{j+1}P_{j+1}^{2a+1,2b+1}(x)
                                         -B_{j-1}P_{j-1}^{2a+1,2b+1}(x)\},
\end{equation}
we get
\begin{eqnarray}
{(g^{(4)}_{j})}^{1/2}\psi^{(4)}_{j}(x) &=&
\gamma_{j-1}^{(j-1)}w_{a,b}\left[A_{j-1}P_{j-1}^{2a+1,2b+1}(x)
-B_{j-3}P_{j-3}^{2a+1,2b+1}(x)\right]
+\sum_{k=0}^{j-1} \gamma_{k-1}^{(j-1)}{(g^{(4)}_{k})}^{1/2}\psi^{(4)}_{k}(x)\\
{(g^{(1)}_{j})}^{1/2}\phi^{(1)}_{j}(x) &=&
\gamma_{j}^{(j)}w_{a,b}\left[A_{j}P_{j}^{2a+1,2b+1}(x)
-B_{j-2}P_{j-2}^{2a+1,2b+1}(x)\right]+\sum_{k=0}^{j-1}
\gamma_{k}^{(j)}{(g^{(1)}_{k})}^{1/2}\phi^{(1)}_{k}(x).
\end{eqnarray}

In this paper we will give the proof for $\beta=4$. The proof for
$\beta=1$ follows the same line of logic and can be found in
\cite{ghosh} and \cite{ghoshpandey}.

Using (\ref{wab1}) and orthonormality of $P^{2a+1,2b+1}_{j}(x)$
w.r.t. the weight function $w_{2a+1,2b+1}(x)$, the scalar products

\begin{eqnarray}
\left(\phi^{(4)}_{2m}(x),\psi^{(4)}_{2m-2k}(x)\right)=0,\Longrightarrow
\gamma_{2m-2k}^{(2m-1)}=0,\qquad k=1,2,\ldots.
\end{eqnarray}
We also have
\begin{eqnarray}
\left(\phi^{(4)}_{2m}(x),\psi^{(4)}_{2m-2k+1}(x)\right) &=& 0,\Longrightarrow
\gamma_{2m-2k-1}^{(2m-1)}=0,\qquad k=2,3,\ldots,\\
\left(\phi^{(4)}_{2m+1}(x),\psi^{(4)}_{2m-2k}(x)\right) &=& 0,\Longrightarrow
\gamma_{2m-2k}^{(2m)}=0,\qquad k=1,2,\ldots,\\
\left(\phi^{(4)}_{2m+1}(x),\psi^{(4)}_{2m-2k+1}(x)\right)=0,\Longrightarrow
\gamma_{2m-2k-1}^{(2m)}=0,\qquad k=1,2, \ldots.
\end{eqnarray}

Since odd SOP is arbitrary to the addition of any multiple of the
lower even SOP, we set $\gamma_{2m-1}^{(2m)}=0$.
Choosing
$\gamma_{2m-1}^{(2m-1)}=-\frac{1}{A_{2m-1}}$ and
$\gamma_{2m}^{(2m)}=1$,
we get
\begin{eqnarray}
g^{(4)}_{2m}=g^{(4)}_{2m+1}=h^{2a+1,2b+1}_{2m-1}.
\end{eqnarray}
Here we have used (\ref{abh}). Finally,
\begin{eqnarray}
\left(\psi^{(4)}_{2m}(x),\phi^{(4)}_{2m-1}(x)\right)=0,\Longrightarrow
\gamma_{2m-3}^{(2m-1)}=\frac{B_{2m-3}}{A_{2m-1}},\qquad m\geq 2.
\end{eqnarray}
Thus we get (\ref{dual41}-\ref{duality42}).

To prove (\ref{dual43}-\ref{dual44}), we start with the expansion
\begin{eqnarray}
{(g^{(4)}_{j})}^{1/2}\phi^{(4)}_{j}(x) &=&
\gamma_{j-1}^{(j-1)}2^{a+3/2}w_{a+1}(x)L_{j-1}^{2a+1}(2x)+\sum_{k=0}^{j-1}
\gamma_{k-1}^{(j-1)}{(g^{(4)}_{k})}^{1/2}\phi^{(4)}_{k}(x), \qquad
j\geq 1,\qquad
\gamma_{j-1}^{(j-1)}\neq 0,\\
{(g^{(1)}_{j})}^{1/2}\psi^{(1)}_{j}(x) &=&
\gamma_{j}^{(j)}2^{a+3/2}w_{a+1}(x)L_{j-1}^{2a+1}(2x)+\sum_{k=0}^{j-1}
\gamma_{k}^{(j)}{(g^{(1)}_{k})}^{1/2}\psi^{(1)}_{k}(x),\qquad j\geq
1,\qquad \gamma_{j}^{(j)}\neq 0.
\end{eqnarray}
Using (\ref{derivativel}), we get $\psi^{(4)}_{j}(x)$ and
$\phi^{(1)}_{j}(x)$. Finally, we follow the same procedure and use
(\ref{gammal})-(\ref{abhL}) to obtain
(\ref{dual43})-(\ref{duality45}).

For Gaussian SOP, we expand
\begin{eqnarray}
{(g^{(4)}_{j})}^{1/2}\phi^{(4)}_{j}(x) &=&
\gamma_{j-1}^{(j-1)}w(x)H_{j}(x)+\sum_{k=0}^{j-1}
\gamma_{k-1}^{(j-1)}{(g^{(4)}_{k})}^{1/2}\phi^{(4)}_{k}(x), \qquad
j=0,1,\ldots,\qquad
\gamma_{j-1}^{(j-1)}\neq 0,\\
{(g^{(1)}_{j})}^{1/2}\psi^{(1)}_{j}(x) &=&
\gamma_{j}^{(j)}w(x)H_{j-1}(x)+\sum_{k=0}^{j-1}
\gamma_{k}^{(j)}{(g^{(1)}_{k})}^{1/2}\psi^{(1)}_{k}(x),\qquad j\geq
1,\qquad \gamma_{j}^{(j)}\neq 0,
\end{eqnarray}
and use the relation
\begin{eqnarray}
\frac{d}{dx}\left(e^{-x^2/2}H_{j}(x)\right)=e^{-x^2/2}\left(-\frac{1}{2}H_{j+1}(x)+jH_{j-1}(x)\right)
\end{eqnarray}
to prove (\ref{dual45}-\ref{dual46}).

\section{Classical orthogonal polynomials and some relevant formula}

Orthogonal polynomials $P_{j}(x)$ of order $j$, associated with
weight function $w(x)$ in the interval $[x_{1},x_{2}]$ is defined as
\cite{szego}:
\begin{eqnarray}
\int_{x_{1}}^{x_{2}}P_{j}(x)P_{k}(x)w(x)dx=h_{j}\delta_{j,k},\qquad
P_{j}(x)=\sum_{l=0}^{j}k_{l}^{(j)}x^{l}, \qquad j,k,l\in\mathbb N,
\end{eqnarray}
where $h_{j}$ is the normalization constant and $k_{j}^{(j)}$ is the
leading coefficient. They satisfy three-term recursion relation
\begin{eqnarray}
\label{recursionq1}
xP_{j}(x)=Q_{j,j+1}P_{j+1}(x)+Q_{j,j}P_{j}(x)+Q_{j,j-1}P_{j-1}(x),\qquad
j=0,1, \ldots
\end{eqnarray}
where $Q_{j,k}$ is the recursion coefficient.

For classical weight functions (\ref{jacobi}, \ref{laguerre},
\ref{gaussian}), defined in the interval $[x_{1},x_{2}]$, we have
\cite{szego}:

\subsection*{\bf Table 1 }

\begin{tabular}{|c|c|c|c|}
\hline
& Jacobi weight: $[-1,1]$ &Associated  Laguerre weight: $[0,\infty]$ & Gaussian weight: $[-\infty,\infty]$\\
\hline
$P_{0}(x)$& 1 & 1 & 1 \\
\hline
$P_{1}(x)$& $ \frac{1}{2}(a+b+2)x+\frac{1}{2}(a-b)$ & $-x+a+1$ & $2x$ \\
\hline ($h_{j}$) &
$h_{j}^{a,b}=\left(\frac{2^{a+b+1}}{(2j+a+b+1)}\frac{\Gamma(j+a+1)\Gamma(j+b+1)}
                                   {\Gamma(j+1)\Gamma(j+a+b+1)}\right)$ & $ h_{j}^{a}=\frac{\Gamma(j+a+1)}{j!}$
                                   & $h_{j}=\pi^{1/2}2^{j}j!$ \\
\hline $k^{(j)}_{j}$&  $k_{j}^{a,b}=\frac{1}{2^j}
                     \left(\begin{tabular}{clcr}2j+a+b\\j\end{tabular}\right)$.
 & $k_{j}^{a}=\frac{{(-1)}^j}{j!}$ & $2^{j}$ \\
\hline $Q_{j-1,j}$ &
$\frac{2j(j+a+b)}{(2j+a+b)(2j+a+b-1)}$ & $ -j$ & 1/2 \\
\hline
$Q_{j-1,j-1}$ & $\frac{(b^2-a^2)}{(2j+a+b)(2j+a+b-2)}$ & $2j+a-1$ & 0 \\
\hline $Q_{j-1,j-2}$ &  $\frac{2(j+a-1)(j+b-1)}{(2j+a+b-1)(2j+a+b-2)}$ & $-j+a-1$ & $j-1$ \\
\hline
\end{tabular}


\subsection{Asymptotic formula}

Here we give a brief summary of the asymptotic results \cite{szego}
of OP with classical weight which will be useful
in our analysis of the corresponding SOP.

Jacobi polynomial $P^{(a,b)}_{j}(x)$, for large $j$,  arbitrary real
$a,b$ and fixed positive number $\epsilon$, is written as
\begin{eqnarray}
\nonumber
&& x=\cos\theta,\qquad \epsilon \leq \theta\leq \pi-\epsilon,\\
\label{asymjac2}
&&{({h}_{j}^{a,b})}^{-1/2}
      {(w_{a,b}(x))}^{1/2}
              P_{j}^{(a,b)}(x)
                                 =\sqrt{\frac{2}{\pi\sin\theta}}
        \cos[(j+\frac{a+b+1}{2})\theta-(a+\frac{1}{2})\frac{\pi}{2}]+O(j^{-1}),
\end{eqnarray}
where we have used \cite{szego} and ${h}_{j}^{a,b}\simeq
2^{(a+b)}j^{-1}$.

Associated Laguerre polynomial $L^{a}_{j}(x)$, for large $j$, has
formula of the Plancherel-Rotach type \cite{szego} given below. For
`$a$' arbitrary and real, $\epsilon$ a fixed positive number, we
have for

\begin{eqnarray}
\nonumber && \noindent x = (4j+2a+2)\cos^{2}\theta, \qquad \epsilon
\leq \theta \leq
\pi/2-\epsilon j^{-\frac{1}{2}},\\
\label{asylag2}
 && e^{-x/2}L_{j}^{(a)}(x) =
                         \frac{{(-1)}^{j}}{\sqrt{2\pi j\sin\theta\cos\theta}}
{\left(\frac{j}{x}\right)}^{a/2}\left\{\sin[(j+(a+1)/2)(\sin2\theta
-2\theta)+\frac{3\pi}{4}]\right\}+{(jx)}^{-\frac{1}{2}}O(1),
\end{eqnarray}
where $h_{j}^{(a)}\simeq j^{a}$. Unlike the Jacobi case, for given
$x$, $\theta$ depends on $j$; for example $\theta_{j}-\theta_{j\mp
1}\simeq \pm {(2j\tan\theta_{j})}^{-1}$. Then  with
$\theta\equiv\theta_{j}$, we can also write
\begin{eqnarray}
\nonumber e^{-x/2}L_{j\mp \Delta j}^{(a)}(x)
&=&\frac{{(-1)}^{j-1}}{\sqrt{2\pi j\sin\theta\cos\theta}}
{\left(\frac{j}{x}\right)}^{a/2}\left\{\sin[(j+(a+1)/2)(\sin2\theta
-2\theta) \pm
2\theta\Delta j+\frac{3\pi}{4}]\right\}+{(jx)}^{-\frac{1}{2}}O(1).\\
\label{asylag2deltaj}
\end{eqnarray}
Equation (\ref{asylag2deltaj}) will be useful in deriving the
asymptotic formul\ae for SOP.

Finally, Hermite polynomial $H_{j}(x)$, for large $j$, has formula
of the Plancherel-Rotach type given below. For $\epsilon$ a fixed
positive number, we have

\begin{eqnarray}
\nonumber && x = {(2j+1)}^{\frac{1}{2}}\cos\theta, \qquad \epsilon
\leq \theta \leq
\pi-\epsilon,\\
&& e^{-x^{2}/2}H_{j}(x)
=\frac{{(h_{j})}^{1/2}}{\sqrt{\pi\sin\theta}}{\left({\frac{2}{j}}\right)}^{1/4}
\sin[(j/2+1/4)(\sin2\theta -2\theta)+\frac{3\pi}{4}]+O(j^{-1}).
\end{eqnarray}
Unlike the Jacobi case, for given $x$, $\theta$ depends on $j$; for
example $\theta_{j}-\theta_{j\mp 1}\simeq \pm
{(2j\tan\theta_{j})}^{-1}$. Then  with $\theta\equiv\theta_{j}$, we
can also write
\begin{equation}
e^{-x^{2}/2}H_{j\mp 1}(x) =\frac{{(h_{j\mp
1})}^{1/2}}{\sqrt{\pi\sin\theta}}{\left({\frac{2}{j}}\right)}^{1/4}
\sin[(j/2+1/4)(\sin2\theta -2\theta)
\pm\theta+\frac{3\pi}{4}]+O(j^{-1}),
\end{equation}
where we have used again $\theta_{j}-\theta_{j\mp
1}\simeq\pm{(2j\tan\theta_{j})}^{-1}$.


\section{Universality for orthogonal ensembles}

In this section, we use GCD formula for orthogonal ensembles of
random matrices (\ref{rgcd1}) along with the asymptotic results of
SOP \cite{ghosh,ghoshpandey} to prove that
with proper scaling, the kernel function
$(S^{(1)}_{2N}(x,y)/S^{(1)}_{2N}(x,x))$ and hence the correlation
function for Jacobi orthogonal and associated Laguerre orthogonal
ensemble is stationary and universal.

\subsection{Jacobi SOP}

The asymptotic results for the SOP are derived
using (\ref{asymjac2}) in (\ref{psijac1}-\ref{phijac1}). Here,
$A_{j}\approx B_{j} \approx -j/2$ and $\gamma_{j}\approx 1$ for
large $j$. For
\begin{eqnarray}
&& \nonumber x=\cos\theta,\qquad \epsilon \leq \theta\leq \pi-\epsilon,\\
&& \label{asymjaclag1psiodd} \psi^{(1)}_{2m+1}(x)
          = \sqrt{\frac{2\sin\theta}{\pi}}
                                \cos[(2m+a+b+\frac{3}{2})\theta-(2a+\frac{3}{2})
                                \frac{\pi}{2}]+O{(2m)}^{-1},\\
\label{asymjac1} && \psi^{(1)}_{2m}(x)
          =-\frac{1}{m\sqrt{2\pi \sin\theta}}\left[\sin[(2m+a+b+\frac{3}{2})\theta-(2a+\frac{3}{2})
                                \frac{\pi}{2}]+O(1)+O{(2m)}^{-1}\right].
\end{eqnarray}
\begin{eqnarray}
&& \label{asymjaclag1phiodd} \phi^{(1)}_{2m+1}(x)
          = 2m\sqrt{\frac{2}{\pi\sin\theta}}
                                \left[\sin[(2m+a+b+\frac{3}{2})\theta-(2a+\frac{3}{2})
                                \frac{\pi}{2}]+O{(2m)}^{-1}\right],\\
\label{asymjac1phi} && \phi^{(1)}_{2m}(x)
          =\sqrt{\frac{2}{\pi\sin^{3}\theta}}
                                \cos[(2m+a+b+\frac{3}{2})\theta-(2a+\frac{3}{2})
                                \frac{\pi}{2}]+O{(2m)}^{-1}.
\end{eqnarray}
Equations (\ref{asymjaclag1psiodd}), (\ref{asymjaclag1phiodd}) and
(\ref{asymjac1phi}) are obtained by using (\ref{asymjac2}) in
(\ref{psijac1}), (\ref{phijac1odd}) and (\ref{phijac1}). Equation
(\ref{asymjac1}) is obtained by partial integration of
(\ref{asymjac1phi}).  Here, we note that by directly differentiating
$\psi^{(1)}(x)$, we can get $\phi^{(1)}(x)$ to the leading order,
thereby confirming our result.

\subsection{Universality in Jacobi orthogonal ensemble}

In this subsection, we calculate the level density of Jacobi
orthogonal ensemble and show that in the bulk of the spectrum, the
scaled or ``unfolded'' \cite{mehta} kernel function
$(S^{(1)}_{2N}(x,y)/S^{(4)}_{2N}(x,x))$ is stationary and universal.

To study the kernel function (\ref{rgcd1}), we calculate
$P^{(1)}_{2N-1,2N}$, $R^{(1)}_{2N-1,2N+1}$, $R^{(1)}_{2N-1,2N}$,
$R^{(1)}_{2N-2,2N}$. We expand
\begin{eqnarray}
(1-x^2)\phi^{(1)}_{2m+1}(x)=\sum^{2m+2}_{j=2m-2}P^{(1)}_{2m+1,j}\psi^{(1)}_{j}(x),
\end{eqnarray}
\begin{eqnarray}
x(1-x^2)\phi^{(1)}_{2m+1}=\sum^{2m+3}_{j=2m-2}R^{(1)}_{2m+1,j}\psi^{(1)}_{j}(x),
\qquad
x(1-x^2)\phi^{(1)}_{2m}(x)=\sum^{2m+2}_{2m-2}R^{(1)}_{2m,j}\psi^{(1)}_{j}(x).
\end{eqnarray}
We get

\begin{eqnarray}
P^{(1)}_{2m+1,2m+2}=\sqrt{\frac{g^{(1)}_{2m+2}}{g^{(1)}_{2m}}}
\frac{(2m+1)(2m+2)(2m+2a+2b+3)(2m+2a+2b+4)}{(4m+2a+2b+3)(4m+2a+2b+5)},
\end{eqnarray}
\begin{eqnarray}
R^{(1)}_{2m+1,2m+3}=-2\sqrt{\frac{g^{(1)}_{2m+2}}{g^{(1)}_{2m}}}
\frac{(2m+1)(2m+2)(2m+2a+2b+3)(2m+2a+2b+4)}{(4m+2a+2b+3)(4m+2a+2b+5)(4m+2a+2b+6)},
\end{eqnarray}
\begin{eqnarray}
\nonumber R^{(1)}_{2m+1,2m+2}&=&
\sqrt{\frac{g^{(1)}_{2m+2}}{g^{(1)}_{2m}}}
\frac{[{(2b+1)}^2-{(2a+1)}^2](2m+1)(2m+2)(2m+2a+2b+3)(2m+2a+2b+4)}{(4m+2a+2b+3)(4m+2a+2b+4)(4m+2a+2b+5){(4m+2a+2b+6)}},\\
&=& 0, \qquad{\rm for}\qquad a=b,
\end{eqnarray}
\begin{eqnarray}
R^{(1)}_{2m,2m+2}=-2\sqrt{\frac{g^{(1)}_{2m+2}}{g^{(1)}_{2m}}}
\frac{(2m+1)(2m+2)(2m+2a+2b+3)(2m+2a+2b+4)}{(4m+2a+2b+3)(4m+2a+2b+4)(4m+2a+2b+5)}.
\end{eqnarray}
For $m=N-1$, large $N$, we have $(g^{(4)}_{2N}/g^{(4)}_{2N-2})\simeq
1$ and
\begin{eqnarray}
\nonumber
&&\qquad P^{(1)}_{2N-1,2N}\sim N^{2}+O(N),\\
\label{approxPR1}
 && R^{(1)}_{2N-1,2N+1}\sim
-\frac{N}{2}+O(1),\qquad R^{(1)}_{2N-1,2N}\sim (b-a)O(1),\qquad
R^{(1)}_{2N-2,2N}\sim -\frac{N}{2}+O(1).
\end{eqnarray}
Finally using (\ref{asymjaclag1psiodd}) and (\ref{asymjac1}),
defined in the $\theta$ interval $[\epsilon,\pi-\epsilon] $, and
(\ref{approxPR1}) in the GCD formula (\ref{rgcd1}), we get for
\begin{eqnarray}
\nonumber  x=\cos\theta,\qquad y=x+\Delta x &=&
\cos(\theta+\Delta\theta),\qquad x-y \simeq
\Delta\theta\sin\theta, \\
\nonumber (x-y)(1-x^2)S^{(1)}_{2N}(x,y) &=&
R^{(1)}_{2N-2,2N}\left[\psi^{(1)}_{2N}(x)\psi^{(1)}_{2N-1}(y)
-\psi^{(1)}_{2N}(y)\psi^{(1)}_{2N-1}(x)\right]\\
\nonumber && +
R^{(1)}_{2N-1,2N+1}\left[\psi^{(1)}_{2N-2}(x)\psi^{(1)}_{2N+1}(y)
-\psi^{(1)}_{2N-2}(y)\psi^{(1)}_{2N+1}(x)\right]\\
\nonumber && + \left(R^{(1)}_{2N-1,2N}-xP^{(1)}_{2N-1,2N}\right)
\left[\psi^{(1)}_{2N-2}(x)\psi^{(1)}_{2N}(y)
-\psi^{(1)}_{2N-2}(y)\psi^{(1)}_{2N}(x)\right] \\
\nonumber &=& \frac{1}{2\pi} \left[\sin (f_{2N}(\theta))\cos
(f_{2N-2}(\theta+\Delta\theta)) -\sin
(f_{2N}(\theta+\Delta\theta))\cos
(f_{2N-2}(\theta))\right]\\
\nonumber && + \frac{1}{2\pi} \left[\sin (f_{2N-2}(\theta))\cos
(f_{2N}(\theta+\Delta\theta)) -\sin
(f_{2N-2}(\theta+\Delta\theta))\cos
(f_{2N}(\theta))\right]\\
\nonumber && - \frac{\cos\theta}{2\pi\sin\theta}\left[\sin
(f_{2N-2}(\theta))\sin (f_{2N}(\theta+\Delta\theta)) -\sin
(f_{2N-2}(\theta+\Delta\theta))\sin (f_{2N}(\theta))\right],
\end{eqnarray}
where, in the second step, we have dropped $ O(1/N^2)$ term. Thus we
get
\begin{eqnarray}
\nonumber (x-y)(1-x^2)S^{(1)}_{2N}(x,y) &=&
\frac{1}{2\pi}\left[\sin[f_{2N}(\theta)-(f_{2N-2}(\theta+\Delta\theta)]
-\sin[f_{2N-2}(\theta)-f_{2N}(\theta+\Delta\theta)]\right]\\
&& \nonumber -
\frac{\cos\theta\sin(2\theta)}{2\pi\sin\theta}\left[\sin[\Delta\theta
f'_{2N}(\theta)]\right]\\
\nonumber &=& -\frac{1}{2\pi}\left[\sin[\Delta\theta
f'_{2N}(\theta)-2\theta]
+\sin[\Delta\theta f'_{2N}(\theta)+2\theta\right]\\
\nonumber && +
\frac{\cos\theta\sin(2\theta)}{2\pi\sin\theta}\left[\sin[\Delta\theta
f'_{2N}(\theta)]\right]\\
&\simeq & -\frac{1}{\pi}\sin[2N\Delta\theta]\left[\cos
(2\theta)-\cos^2\theta\right]
\end{eqnarray}
which gives us finally
\begin{eqnarray} S^{(1)}_{2N}(x,y) &=&
\frac{\sin(2N\Delta\theta)}{\pi\Delta\theta\sin\theta},\\
&=& \frac{\sin(2N{(1-x^2)}^{-1/2}\Delta x)}{\pi\Delta x},\qquad
|x|<1.
\end{eqnarray}
With $x\rightarrow y$, we get the level density $\rho(x)$
\begin{eqnarray}
S^{(1)}_{2N}(x,x):=\rho(x) &=& \frac{2N}{\pi\sqrt{1-x^2}},\qquad
|x|<1.
\end{eqnarray}
With $\Delta x\rightarrow 0$, i.e. in the bulk of the spectrum, we
get the ``universal'' sine-kernel
\begin{eqnarray}
\label{sin1} \frac{S^{(1)}_{2N}(x,y)}{S^{(1)}_{2N}(x,x)} &=&
\frac{\sin \pi r}{\pi r},\qquad r=\rho(x)\Delta x.
\end{eqnarray}

\subsection{Associated Laguerre SOP}

To obtain the asymptotic properties of associated Laguerre SOP, we use the relations 
$2^{a+1/2}\sqrt{x}w_{a}(x)=\sqrt{w_{2a+1}(2x)}$ and
$2^{a+1/2}w_{a+1}(x)=\sqrt{x}\sqrt{w_{2a+1}(2x)}$. Replacing this in
(\ref{psi1lagodd})-(\ref{phi1lageven}) and using
(\ref{asylag2}) we get the asymptotic formula.

For arbitrary $a$ and $\epsilon$ a fixed positive number,
\begin{eqnarray}
\nonumber
&& 2x=(8m+4a+4)\cos^{2}\theta,\qquad \epsilon\leq\theta\leq \pi/2-\epsilon m^{-1/2},\qquad \theta\equiv {\theta}_{2m},\\
\label{asyphieven1lag} \phi^{(1)}_{2m}(x)
                &=& \frac{1}{4m\sqrt{\pi\sin\theta\cos^{3}\theta}}
                \left[\sin(f_{2m}(\theta))+\frac{O(1)}{\sqrt{2mx}}\right],\\
\label{asypsiodd1lag}\psi^{(1)}_{2m+1}(x)
                &=& \frac{2}{\sqrt{\pi\tan\theta}}
                  \left[\sin(f_{2m}(\theta))+\frac{O(1)}{\sqrt{2mx}}\right],
\end{eqnarray}
\begin{eqnarray}
\label{asypsieven1lag} \psi^{(1)}_{2m}(x)
                &=&-\frac{1}{4m\sqrt{\pi\sin^{3}\theta\cos\theta}}
                \left[\cos(f_{2m}(\theta))+O(1)+\frac{O(1)}{\sqrt{2mx}}\right],\\
\label{asyphiodd1lag} \phi^{(1)}_{2m+1}(x)
                &=& 2\sqrt{\frac{\tan\theta}{\pi}}
                 \left[\cos(f_{2m}(\theta))+\frac{O(1)}{\sqrt{2mx}}\right],
\end{eqnarray}
with
\begin{eqnarray}
\label{asyphieven1lagdelta} \phi^{(1)}_{2m\pm 2}(x)
                &=& \frac{1}{4m\sqrt{\pi\sin\theta\cos^{3}\theta}}
                \left[\sin(f_{2m}(\theta)\mp4\theta)+\frac{O(1)}{\sqrt{2mx}}\right].
\end{eqnarray}
Similar relations hold for (\ref{asypsiodd1lag}),
(\ref{asypsieven1lag}) and (\ref{asyphiodd1lag}) as $m\rightarrow
m\pm 1$. Here
\begin{eqnarray}
f_{2m}(\theta)=(2m+a+1)(\sin2\theta-2\theta)+\frac{3\pi}{4}.
\end{eqnarray}


\subsection{Universality in associated Laguerre orthogonal ensemble}

In this subsection, we calculate the level density of associated
Laguerre orthogonal ensemble and show that in the bulk of the
spectrum, the scaled or ``unfolded'' \cite{mehta} kernel function
$(S^{(1)}_{2N}(x,y)/S^{(1)}_{2N}(x,x))$ is stationary and universal.

To calculate the kernel function (\ref{rgcd1}), we need
$R^{(1)}_{2N-1,2N+1}$, $R^{(1)}_{2N-1,2N}$, $R^{(1)}_{2N-2,2N}$ and
$P^{(1)}_{2N-1,2N}$. For this, we expand
\begin{eqnarray}
\label{plag1}
x\phi^{(1)}_{2m+1}(x)     &=& \sum_{j=2m-2}^{2m+2}P^{(1)}_{2m+1,j}\psi^{(1)}_{j}(x),\\
\label{Rlag1even} x^{2}\phi^{(1)}_{2m}(x)&=&
\sum_{j=2m-2}^{2m+2}R^{(1)}_{2m,j}\psi^{(1)}_{j}(x),\qquad
 x^{2}\phi^{(1)}_{2m+1}(x) =
\sum_{j=2m-2}^{2m+3}R^{(1)}_{2m+1,j}\psi^{(1)}_{j}(x),
\end{eqnarray}
Using (\ref{phi1lagodd}) in (\ref{plag1}), we get
\begin{eqnarray}
P^{(1)}_{2m+1,2m+2}=\frac{1}{2}\sqrt{\frac{g^{(1)}_{2m+2}}{g^{(1)}_{2m}}}(2m+1)(2m+2).
\end{eqnarray}
Using (\ref{phi1lageven}) and (\ref{phi1lagodd}) in
(\ref{Rlag1even}), we get
\begin{eqnarray}
R^{(1)}_{2m,2m+2}=R^{(1)}_{2m+1,2m+3}=-\frac{1}{2}\sqrt{\frac{g^{(1)}_{2m+2}}{g^{(1)}_{2m}}}(m+1)(2m+1),
\end{eqnarray}
and
\begin{eqnarray}
R^{(1)}_{2m+1,2m+2} &=&
\frac{1}{2}\sqrt{\frac{g^{(1)}_{2m+2}}{g^{(1)}_{2m}}}(m+1)(2m+1)(4m+2a+4).
\end{eqnarray}

For $m=N-1$, large $N$, we have $(g^{(1)}_{2N}/g^{(1)}_{2N-2})\simeq
1$ and
\begin{eqnarray}
\nonumber
&&\qquad P^{(1)}_{2N-1,2N}\sim 2N^{2}+O(N),\\
\label{approxPR1lag}
 && R^{(1)}_{2N-1,2N+1}\sim
-N^{2}+O(N),\qquad R^{(1)}_{2N-1,2N}\sim 4N^{3}+O(N^2),\qquad
R^{(1)}_{2N-2,2N}\sim -N^{2}+O(N).
\end{eqnarray}
Finally using (\ref{asypsiodd1lag}), (\ref{asypsieven1lag}) and
(\ref{asyphieven1lagdelta}), defined in the $\theta$ interval
$[\epsilon,\pi/2-\epsilon N^{-1/2}] $, and (\ref{approxPR1lag}) in
the GCD formula (\ref{rgcd1}), we get for
\begin{eqnarray}
\nonumber  x &=& (4N+2a+2)\cos^{2}\theta, \qquad y=x+\Delta x, \\
\nonumber x(x-y)S^{(1)}_{2N}(x,y) &=&
R^{(1)}_{2N-2,2N}\left[\psi^{(1)}_{2N}(x)\psi^{(1)}_{2N-1}(y)
-\psi^{(1)}_{2N}(y)\psi^{(1)}_{2N-1}(x)\right]\\
\nonumber &+&
R^{(1)}_{2N-1,2N+1}\left[\psi^{(1)}_{2N-2}(x)\psi^{(1)}_{2N+1}(y)
-\psi^{(1)}_{2N-2}(y)\psi^{(1)}_{2N+1}(x)\right]\\
\nonumber &+&
(R^{(1)}_{2N-1,2N}-xP^{(1)}_{2N-1,2N})\left[\psi^{(1)}_{2N-2}(x)\psi^{(1)}_{2N}(y)
-\psi^{(1)}_{2N-2}(y)\psi^{(1)}_{2N}(x)\right] \\
\nonumber &=& \frac{N}{2\pi\sin^2\theta} \left[\sin
(f_{2N-2}(\theta+\Delta\theta)-f_{2N}(\theta))+\sin
(f_{2N}(\theta+\Delta\theta)-(f_{2N-2}(\theta)))\right]\\
\nonumber && -\left[\frac{N}{2\pi\sin^2\theta\tan 2\theta}\right]
[\cos (f_{2N-2}(\theta))\cos (f_{2N}(\theta+\Delta\theta))
 -\cos(f_{2N-2}(\theta+\Delta\theta))\cos (f_{2N}(\theta))] \\
 &=&\frac{N\cos 4\theta}{\pi\sin^2\theta} \sin\left(
\Delta\theta\frac{\partial f_{2N}(\theta)}{\partial\theta}\right)
-\left[\frac{N\sin 4\theta}{2\pi\sin^2\theta\tan 2\theta}\right]
\sin \left(\Delta\theta\frac{\partial
f_{2N}(\theta)}{\partial\theta}\right).
\end{eqnarray}
Here, we have neglected the oscillatory term of $O(1)$ arising from
the even function. Finally, we get
\begin{eqnarray} \nonumber (x-y)S^{(1)}_{2N}(x,y)
&=& \frac{1}{\pi\sin^{2}2\theta}(\cos 4\theta-\cos^{2} 2\theta)
\left[\sin \left(\Delta\theta\frac{\partial
f_{2N}(\theta)}{\partial\theta}\right)\right]\\
 &=& \frac{1}{\pi}\sin[8N\sin^2\theta\Delta\theta].
\end{eqnarray}
Combining all, we get
\begin{eqnarray}
S^{(1)}_{2N}(x,y) = \frac{\sin(x^{-1/2}\sqrt{(4N-x)}\Delta
x)}{\pi\Delta x},\qquad 0<x< 4N.
\end{eqnarray}
With $ x\rightarrow y$, we get the level density
\begin{eqnarray}
S^{(1)}_{2N}(x,x) &=& \frac{1}{\pi}\sqrt{\frac{4N-x}{x}},\qquad 0<x<
4N.
\end{eqnarray}
 With
$\Delta x\rightarrow 0$, (i.e. in the bulk of the spectrum) and
$r=\Delta xS^{(1)}_{2N}(x,x)$, we get the ``universal'' sine-kernel
(\ref{sin1}).



\section{Universality for symplectic ensembles}

Before we prove  universality of the eigenvalue correlation for
symplectic ensembles, we would clarify some of the confusing
notations related to the corresponding SOP.

SOP corresponding to symplectic ensembles of
random matrices can be defined in an interval $[x_{1},x_{2}]$ as

\begin{eqnarray}
\label{def1}
 \int_{x_1}^{x_2}{g_{j}}^{-1}\left[\pi^{(4)}_{j}(x){\pi^{(4)}_{k}}'(x)-\pi^{(4)}_{k}(x){\pi^{(4)}_{j}}'(x)\right]w(x)dx
=
\int_{x_1}^{x_2}\left[\phi^{(4)}_{j}(x)\psi^{(4)}_{k}(x)-\phi^{(4)}_{k}(x)\psi^{(4)}_{j}(x)\right]dx
= Z_{j,k},
\end{eqnarray}
where
\begin{eqnarray}
\label{skeworthogonal4a}
\phi^{(4)}_{j}(x)={(g_{j})}^{-\frac{1}{2}}{w(x)}^{1/2}\pi^{(4)}_{j}(x),\qquad
\psi^{(4)}_{j}(x)=\frac{d}{dx}\phi^{(4)}_{j}(x).
\end{eqnarray}
Here, $\pi^{(4)}_{j}(x)$ are SOP defined with respect to $w(x)$.
This definition is used in \cite{ghosh} and \cite{ghoshpandey} and
will also be used in this paper to study the statistical properties
of the symplectic ensembles.

An alternative definition is if we write
\begin{eqnarray}
\phi^{(4)}_{j}(x)={(g_{j})}^{-\frac{1}{2}}w(x)\pi^{(4)}_{j}(x),\qquad
\psi^{(4)}_{j}(x)=\frac{d}{dx}\phi^{(4)}_{j}(x),
\end{eqnarray}
such that
\begin{eqnarray}
\nonumber
\frac{1}{2}\int_{x_1}^{x_2}{g_{j}}^{-1}\left[\pi^{(4)}_{j}(x){\pi^{(4)}_{k}}'(x)-\pi^{(4)}_{k}(x){\pi^{(4)}_{j}}'(x)\right]w^{2}(x)dx
&=&
 \frac{1}{2}\int_{x_1}^{x_2}\left[\phi^{(4)}_{j}(x)\psi^{(4)}_{k}(x)-\phi^{(4)}_{k}(x)\psi^{(4)}_{j}(x)\right]dx\\
\nonumber
&=& \int_{x_1}^{x_2}\phi^{(4)}_{j}(x)\psi^{(4)}_{k}(x)dx \\
\label{def2} &=& Z_{j,k}.
\end{eqnarray}
Here \label{skeworthogonal4b}
$\phi^{(4)}_{j}(x_{1})=\phi^{(4)}_{j}(x_{2})=0$. The SOP
$\pi^{(4)}_{j}(x)$ in this definition is defined with respect to
$w^{2}(x)$. This definition is used in \cite{ghosh3} and
\cite{ghosh4} and is used to prove duality between SOP
 corresponding to orthogonal and symplectic ensembles of
random matrices. Also, this definition differs from that of
\cite{mehta,ghosh,ghoshpandey} for $\beta=4$ by a factor $2$, which
is incorporated in the normalization constant.

We will use (\ref{def1}) and the GCD formula to prove universality.
Here we would like to mention that our GCD results are valid for
both these definitions with some minor difference in ${\overline
R}^{(4)}(x)$.

\subsection{Jacobi SOP}

We consider  SOP (\ref{skeworthogonal4a})
defined with respect to the Jacobi weight (\ref{jacobi}). As shown
in \cite{ghosh} and \cite{ghoshpandey},
$\pi^{(4)}_{j}(x)\equiv\pi_{j}(x)$ and $\pi_{j}^{\prime} (x)$ can be
written compactly in terms of Jacobi OP
$P_{j}^{a,b}(x)$:
\begin{eqnarray}
\label{oddpiprime}
{\pi}_{2m+1}^{\prime} (x) &=&P_{2m}^{a,b}(x),\qquad m=0,1,\ldots,\\
\label{evenpiprime} \pi_{2m}^{\prime} (x)
&=&P_{2m-1}^{a,b}(x)+\eta_{2m}\pi_{2m-2}^{\prime}(x),\qquad
m=1,2,\ldots,\qquad \pi_{0}^{\prime} (x)=0,
\end{eqnarray}
where $\eta_{2m}$ is a constant, given in Eq.(\ref{F}) below. On
integration, we find the polynomials:
\begin{eqnarray}
\label{oddpi} \pi_{2m+1}(x)
&=&\frac{2}{(2m+a+b)}\left[D_{2m+1}P_{2m+1}^{a,b}(x)+
             E_{2m+1}P_{2m}^{a,b}(x)+F_{2m+1}P_{2m-1}^{a,b}(x)\right],\qquad
m=0,1,\ldots,\\
\nonumber
\pi_{2m}(x) &=&\frac{2}{(2m+a+b-1)}\\
\label{evenpi}
&&\times\left[D_{2m}P_{2m}^{a,b}(x)+
             E_{2m}P_{2m-1}^{a,b}(x)+F_{2m}P_{2m-2}^{a,b}(x)\right]
                 +\eta_{2m}{\pi}_{2m-2}(x),\qquad
m=0,1,\ldots.
\end{eqnarray}
Here, $P_{j}^{a,b}(x)=0$ for negetive $j$. In (\ref{oddpi}) and
(\ref{evenpi}) we have used the indefinite integral
\begin{eqnarray}
\nonumber
\frac{1}{2}(j+a+b)\int P_{j}^{a,b}(x)dx &=&P_{j+1}^{a-1,b-1}(x),\\
                                        &=&D_{j+1}P_{j+1}^{a,b}(x)+
                                           E_{j+1}P_{j}^{a,b}(x)+
                                           F_{j+1} P_{j-1}^{a,b}(x).
\end{eqnarray}
The integration constants  have been put equal to zero because of
skew-orthogonality with $\pi_{1}(x)$. The constants $D_{j}$,
$E_{j}$, $F_{j}$, $\eta_{j}$ and $g^{(4)}_{j}$ are given by
\begin{eqnarray}
\nonumber D_{j}
&=&\frac{(j+a+b)(j+a+b-1)}{(2j+a+b)(2j+a+b-1)},\qquad
E_{j}          =\frac{(a-b)(j+a+b-1)}{(2j+a+b)(2j+a+b-2)},\\
\label{F} F_{j}
&=&-\frac{(j+a-1)(j+b-1)}{(2j+a+b-1)(2j+a+b-2)},\qquad
\eta_{j}       =\frac{(j+a-1)(j+b-1)(2j+a+b-5)}{(j-1)(j+a+b-1)(2j+a+b-1)},\\
\nonumber g^{(4)}_{2m}=g^{(4)}_{2m+1} &=& \frac{2h_{2m}^{a,b}}{4m+a+b-1},\\
               &=& \frac{2^{a+b+2}\Gamma(2m+a+1)\Gamma(2m+b+1)}
                  {(4m+a+b+1)(4m+a+b-1)\Gamma(2m+1)\Gamma(2m+a+b+1)}.
\end{eqnarray}
For large $j$ and large $m$,
\begin{eqnarray}
D_{j}   &=& -F_{j}=\frac{1}{4}+O(j^{-1}), \hspace{1cm} E_{j}=(a-b)\left[\frac{1}{4j}+O(j^{-2})\right],\\
\eta_{j}&=& 1+O(j^{-1}),\hspace{2cm}
g^{(4)}_{2m}=\frac{2^{a+b}}{4m^{2}}+O(m^{-3}),
\end{eqnarray}
and in the same approximation,
\begin{eqnarray}
\label{oddpiapprox}
\pi_{2m+1}(x)=\frac{1}{4m}[P_{2m+1}^{a,b}(x)-P_{2m-1}^{a,b}(x)],\qquad
 \pi_{2m}(x)  =\frac{1}{4m}[P_{2m}^{a,b}(x)
                          +2^{(a+b)/2}{(w_{a,b}(x))}^{-1/2}].
\end{eqnarray}
Here in (\ref{oddpiapprox}), the non-polynomial term on the right
hand side is the large-m approximation for the lower-order terms in
the series in (\ref{evenpi}) and has been verified numerically.

For large $m$, we use (\ref{oddpiapprox}) and the asymptotic formula
for Jacobi OP (\ref{asymjac2}) to obtain
asymptotic formula for Jacobi SOP. For $a$,
$b$ arbitrary and real, $\epsilon$ a fixed positive number, we have
for

\begin{eqnarray}
\nonumber && x = \cos\theta, \qquad \epsilon \leq \theta \leq
\pi-\epsilon ,\\
\label{asymjac4odd} &&
{(g^{(4)}_{2m})}^{-1/2}\sqrt{w_{a,b}(x)}\pi_{2m+1}(x):=\phi^{(4)}_{2m+1}(x)
=-{\sqrt\frac{\sin\theta}{\pi m}}
                                    \sin[f_{2m}(\theta)]+O(m^{-\frac{3}{2}}),\\
\label{asymjac4even} &&
{(g^{(4)}_{2m})}^{-1/2}\sqrt{w_{a,b}(x)}\pi_{2m}(x):=\phi^{(4)}_{2m}(x)
=\frac{1}{2}\left[\frac{1}{\sqrt{\pi m\sin\theta}}
                                  \cos[f_{2m}(\theta)]+1\right]+O(m^{-\frac{3}{2}}),\\
\label{asymjac4oddprime} &&
\psi^{(4)}_{2m+1}(x)
=2{\sqrt\frac{m}{\pi\sin\theta}}
                                    \cos[f_{2m}(\theta)]+O(m^{-\frac{1}{2}}),\\
\label{asymjac4evenprime} &&
\psi^{(4)}_{2m}(x)=\sqrt{\frac{m}{\pi\sin^3\theta}}\left[
                                  \sin[f_{2m}(\theta)]+1\right]+O(m^{-\frac{1}{2}}).
\end{eqnarray}
where
\begin{eqnarray}
f_{2m}(\theta)=\left(2m+\frac{a+b+1}{2}\right)\theta
                                       -\left(a+\frac{1}{2}\right)\frac{\pi}{2}.
\end{eqnarray}

 Equation (\ref{asymjac4oddprime}) is derived from
(\ref{oddpiprime}) and (\ref{asymjac4evenprime}) is obtained by
differentiating (\ref{asymjac4even}). Here, we would like to mention
that to calculate level density and two-point correlation function
for the Jacobi symplectic ensemble, (\ref{asymjac4oddprime}) and
(\ref{asymjac4evenprime}) are not needed. However, they are
important to define the SOP and hence included
for completeness.


\subsection{Level density and 2-point correlation for Jacobi symplectic ensemble}

In this subsection, we calculate the level density of Jacobi
symplectic ensemble and show that in the bulk of the spectrum, the
scaled or ``unfolded'' \cite{mehta} kernel function
$(S^{(4)}_{2N}(x,y)/S^{(4)}_{2N}(y,y))$ is stationary and universal.

As suggested in Eq.(\ref{rgcd4}), to obtain the kernel function, we
need to calculate $R^{(4)}_{2N-1,2N+1}$, $R^{(4)}_{2N-1,2N}$,
$R^{(4)}_{2N-2,2N}$ and $P^{(4)}_{2N-1,2N}$. For this, we use
(\ref{p4Jquatelem})
\begin{eqnarray}
(1-x^2)\frac{d}{dx}\phi^{(4)}_{2m+1}(x)=\sum^{2m+2}_{j=2m-2}P^{(4)}_{2m+1,j}\phi^{(4)}_{j}(x),
\end{eqnarray}
to get $P^{(4)}_{2m+1,2m+2}$. It is given by

\begin{eqnarray}
P^{(4)}_{2m+1,2m+2}=-\sqrt{\frac{g^{(4)}_{2m+2}}{g^{(4)}_{2m}}}\frac{(2m+a+b+1)Q_{2m,2m+1}Q_{2m+1,2m+2}}{2D_{2m+2}},
\end{eqnarray}
where $Q_{j,k}$ and $D_{j}$ are given in the table  and (\ref{F})
respectively.

Similarly, we use (\ref{r4Jquatelem}):
\begin{eqnarray}
x(1-x^2)\frac{d}{dx}\phi^{(4)}_{2m}(x)=\sum^{2m+2}_{j=2m-2}R^{(4)}_{2m,j}\phi^{(4)}_{j}(x),\qquad
x(1-x^2)\frac{d}{dx}\phi^{(4)}_{2m+1}(x)=\sum^{2m+3}_{j=2m-2}R^{(4)}_{2m+1,j}\phi^{(4)}_{j}(x),
\end{eqnarray}
to get $R^{(4)}_{2m,2m+2}$, $R^{(4)}_{2m+1,2m+3}$ and
$R^{(4)}_{2m+1,2m+2}$. For large $m$, we get to the leading order

\begin{eqnarray}
\nonumber R^{(4)}_{2m+1,2m+2} &\simeq&
\sqrt{\frac{g^{(4)}_{2m+2}}{g^{(4)}_{2m}}}
[\frac{(2m+a+b+1)E_{2m+3}Q_{2m,2m+1}Q_{2m+1,2m+2}Q_{2m+2,2m+3}}{2D_{2m+3}D_{2m+2}}\\
&& -(a-b)\frac{(2m+a+b+1)}{2(2m+a+b)}\frac{D_{2m+1}}{D_{2m+2}}Q_{2m+1,2m+2}]+(a-b)O(1/m)\\
&=& 0 \qquad {\rm for}\qquad  a=b,
\end{eqnarray}
\begin{eqnarray}
R^{(4)}_{2m+1,2m+3} \simeq
-\sqrt{\frac{g^{(4)}_{2m+2}}{g^{(4)}_{2m}}}
\frac{(2m+a+b+2)Q_{2m,2m+1}Q_{2m+1,2m+2}Q_{2m+2,2m+3}}{2D_{2m+3}}+O(1),
\end{eqnarray}
\begin{eqnarray}
R^{(4)}_{2m,2m+2} \simeq -\sqrt{\frac{g^{(4)}_{2m+2}}{g^{(4)}_{2m}}}
\frac{(2m+a+b+1)Q_{2m-1,2m}Q_{2m,2m+1}Q_{2m+1,2m+2}}{2D_{2m+2}}+O(1),
\end{eqnarray}
where $Q_{j,k}$, $D_{j}$ and $E_{j}$ are given in the table and
(\ref{F}) respectively.

For large $j$, we use $Q_{j,j+1}\simeq 1/2$. Also for $m=N-1$, large
$N$, we have $(g^{(4)}_{2N}/g^{(4)}_{2N-2})\simeq 1$ and
\begin{eqnarray}
\nonumber
&&\qquad P^{(4)}_{2N-1,2N}\sim -N+O(1),\\
\label{approxPR4}
 && R^{(4)}_{2N-1,2N+1}\sim
-\frac{N}{2}+O(1),\qquad R^{(4)}_{2N-1,2N}\sim
\frac{(b-a)}{4}+(b-a)O(N^{-1}),\qquad   R^{(4)}_{2N-2,2N}\sim
-\frac{N}{2}+O(1).
\end{eqnarray}

Finally using (\ref{asymjac4odd}) and (\ref{asymjac4even}), defined
in the $\theta$ interval $[\epsilon,\pi-\epsilon] $, and
(\ref{approxPR4}) in the GCD formula (\ref{rgcd4}), we get for
\begin{eqnarray}
\nonumber  y=\cos\theta,\qquad x=y+\Delta y &=&
\cos(\theta+\Delta\theta),\qquad y-x \simeq
\Delta\theta\sin\theta, \\
\nonumber (y-x)(1-y^2)S^{(4)}_{2N}(x,y) &=&
R^{(4)}_{2N-2,2N}\left[\phi^{(4)}_{2N}(x)\phi^{(4)}_{2N-1}(y)
-\phi^{(4)}_{2N}(y)\phi^{(4)}_{2N-1}(x)\right]\\
\nonumber && +
R^{(4)}_{2N-1,2N+1}\left[\phi^{(4)}_{2N-2}(x)\phi^{(4)}_{2N+1}(y)
-\phi^{(4)}_{2N-2}(y)\phi^{(4)}_{2N+1}(x)\right]\\
\nonumber && +
\left(R^{(4)}_{2N-1,2N}-xP^{(4)}_{2N-1,2N}\right)\left[\phi^{(4)}_{2N-2}(x)\phi^{(4)}_{2N}(y)
-\phi^{(4)}_{2N-2}(y)\phi^{(4)}_{2N}(x)\right] \\
\nonumber &=& \frac{1}{4\pi}\left[\cos
(f_{2N}(\theta+\Delta\theta))\sin (f_{2N-2}(\theta)) -\cos
(f_{2N}(\theta))\sin
(f_{2N-2}(\theta+\Delta\theta))\right]\\
\nonumber && + \frac{1}{4\pi}\left[\cos
(f_{2N-2}(\theta+\Delta\theta))\sin (f_{2N}(\theta)) -\cos
(f_{2N-2}(\theta))\sin
(f_{2N}(\theta+\Delta\theta))\right]\\
\nonumber && + \frac{\cos\theta}{4\pi\sin\theta}\left[\cos
(f_{2N-2}(\theta+\Delta\theta))\cos (f_{2N}(\theta)) -\cos
(f_{2N-2}(\theta))\cos (f_{2N}(\theta+\Delta\theta))\right] \\
\nonumber &=&
\frac{1}{4\pi}\left[\sin[f_{2N-2}(\theta)-(f_{2N}(\theta+\Delta\theta)]
+\sin[f_{2N}(\theta)-f_{2N-2}(\theta+\Delta\theta)]\right]\\
\nonumber && +
\frac{\cos\theta\sin(2\theta)}{4\pi\sin\theta}\left[\sin[\Delta\theta
f'_{2N}(\theta)]\right]\\
\nonumber &=&
-\frac{1}{4\pi}\left[\sin[(2N+\frac{a+b+1}{2})\Delta\theta-2\theta]
+\sin[(2N+\frac{a+b+1}{2})\Delta\theta+2\theta]\right]\\
\nonumber && +
\frac{\cos\theta\sin(2\theta)}{4\pi\sin\theta}\left[\sin[
(2N+\frac{a+b+1}{2})\Delta\theta]\right]\\
 &\simeq & -\frac{1}{2\pi}\sin[2N\Delta\theta]\left[\cos
(2\theta)-\cos^2\theta\right],
\end{eqnarray}
where in the second step, we have dropped $O(N^{-1})$ terms. This
gives us
\begin{eqnarray} \nonumber
S^{(4)}_{2N}(x,y) &=&
\frac{\sin(2N\Delta\theta)}{2\pi\Delta\theta\sin\theta}, \\
&=& \frac{\sin(2N{(1-y^2)}^{-1/2}\Delta y)}{2\pi\Delta y},\qquad
|y|<1.
\end{eqnarray}
With $\Delta y\rightarrow 0$, we get the level density
\begin{eqnarray}
S^{(4)}_{2N}(y,y) &=& \frac{N}{\pi\sqrt{1-y^2}},\qquad |y|<1.
\end{eqnarray}
In the bulk of the spectrum, we get the ``universal'' sine-kernel
\begin{eqnarray}
\label{sin4} \frac{S^{(4)}_{2N}(x,y)}{S^{(4)}_{2N}(y,y)} &=&
\frac{\sin 2\pi r}{2\pi r},\qquad r=\Delta y S^{(4)}_{2N}(y,y).
\end{eqnarray}

\subsection{Associated Laguerre SOP}

Now we consider SOP (\ref{skeworthogonal4a})
defined with respect to the associated Laguerre weight
(\ref{laguerre}). As shown in \cite{ghosh} and \cite{ghoshpandey},
$\pi^{(4)}_{j}(x)\equiv\pi_{j}(x)$ and $\pi_{j}^{\prime} (x)$ can be
written compactly in terms of associated Laguerre OP $L_{j}^{(a)}(x)$:

\begin{eqnarray}
\label{lag4piprimeodd}
\pi_{2m+1}^{\prime} (x) &=& L_{2m}^{(a)}(x),\qquad m=0,1,\ldots,\\
\label{lag4piprimeeven} \pi_{2m}^{\prime} (x) &=&
L_{2m-1}^{(a)}(x)+{\left(\frac{2m+a-1}{2m-1}\right)}{}\pi_{2m-2}^{\prime}(x),\qquad
m=1,2,\ldots,\qquad \pi_{0}^{\prime} (x)=0.
\end{eqnarray}
 On
integration, we find:
\begin{eqnarray}
\label{lag4piodd}
\pi_{2m+1}(x) &=& -L_{2m+1}^{(a)}(x)+L_{2m}^{(a)}(x),\qquad m=0,1,\ldots,\\
\label{lag4pieven} \pi_{2m}(x) &=&
-L_{2m}^{(a)}(x)+L_{2m-1}^{(a)}(x)+\left(\frac{2m+a-1}{2m-1}\right)\pi_{2m-2},\qquad
m=0,1,\ldots.
\end{eqnarray}

For $a=0$, (\ref{lag4piodd}, \ref{lag4pieven}) give back the results
of \cite{mehta1}, with the observation that any multiple of
$\pi_{2m}(x)$ can be added to $\pi_{2m+1}(x)$. The normalization
constant is given by
\begin{equation}
g^{(4)}_{2m}=g^{(4)}_{2m+1}=-h_{2m}^{(a)}.
\end{equation}
The results (\ref{lag4piodd}, \ref{lag4pieven}) derive from
(\ref{lag4piprimeodd}, \ref{lag4piprimeeven}) from the indefinite
integral,
\begin{eqnarray}
\nonumber \int{\L_{j}^{(a)}}(x)dx =
-L_{j+1}^{(a-1)}(x)=-L_{j+1}^{(a)}(x)+L_{j}^{(a)}(x),
\end{eqnarray}
the constants of integration in (\ref{lag4piodd}),
(\ref{lag4pieven}) being zero on skew-orthogonality with
$\pi_{1}(x)$.

To obtain asymptotic formula for associated Laguerre SOP, we use
(\ref{lag4piodd}),  (\ref{lag4pieven}) and the asymptotic formula
for associated Laguerre OP (\ref{asylag2}) and
(\ref{asylag2deltaj}). To avoid $\theta$ floating inside the
argument, for a given $x$, we choose $\theta$ which effectively
corresponds to $j=2m+1/2$ in (\ref{asylag2}). For `$a$' arbitrary
and real, $\epsilon$ a fixed positive number, we have for

\begin{eqnarray}
\nonumber && x = (8m+2a+4)\cos^{2}\theta, \qquad \epsilon
\leq \theta \leq\pi/2-\epsilon m^{-\frac{1}{2}},\qquad \theta\equiv\theta_{2m+1/2},\\
\label{pievenasylag4} \sqrt{g^{(4)}_{2m}}\phi^{(4)}_{2m}(x)  &=&
-\frac{{(2m)}^{a/2}}{2}\left\{\left[\frac{1}{2\sqrt{\pi
m\cos\theta\sin^{3}\theta}}\cos[f_{2m}(\theta)]+1+\frac{O(1)}{m\sqrt{x}}\right]\right\},\\
\label{pioddasylag4} \sqrt{g^{(4)}_{2m}}\phi^{(4)}_{2m+1}(x)
&=&\frac{{({2m})}^{a/2}}{\sqrt{m\pi\tan\theta}}\left[\sin[f_{2m}(\theta)]+\frac{O(1)}{\sqrt{mx}}\right],\\
 \label{pievenasylag4prime} \sqrt{g^{(4)}_{2m}}\psi^{(4)}_{2m}(x)
&=&{(2m)}^{a/2}\left[\frac{1}{8\sqrt{\pi
m\cos^{3}\theta\sin\theta}}\sin[f_{2m}(\theta)]+\frac{O(1)}{m\sqrt{x}}\right],\\
\label{pioddasylag4prime} \sqrt{g^{(4)}_{2m}}\psi^{(4)}_{2m+1}(x)
&=& {(2m)}^{a/2}\left[\frac{1}{2}\sqrt{\frac{\tan\theta}{m\pi}}
\cos[f_{2m}(\theta)] +\frac{O(1)}{m\sqrt{x}}\right],
\end{eqnarray}
with

\begin{eqnarray}
\label{pievenasylag4delta} \sqrt{g^{(4)}_{2m\pm 2}}\phi^{(4)}_{2m\pm
2}(x) &=& -\frac{{(2m)}^{a/2}}{2}\left\{\left[\frac{1}{2\sqrt{\pi
m\cos\theta\sin^{3}\theta}}\cos[f_{2m}(\theta)\mp
4\theta]+1+\frac{O(1)}{m\sqrt{x}}\right]\right\}.
\end{eqnarray}
Similar relations hold for (\ref{pioddasylag4}),
(\ref{pievenasylag4prime}) and (\ref{pioddasylag4prime}) as
$m\rightarrow m\pm 1$.
Here
\begin{eqnarray}
f_{2m}(\theta)=(2m+1+a/2)(\sin 2\theta-2\theta)+\frac{3\pi}{4}.
\end{eqnarray}
In deriving (\ref{pievenasylag4}) we have used the large-$m$
approximation
\begin{equation}
\left(\frac{8m-x}{2m}\right)\pi_{2m}(x)=
        -\left(L_{2m}^{(a)}(x)+L_{2m+1}^{(a)}(x)\right)
         - \frac{1}{2}\left(\frac{8m-x}{2m}\right)
                            {(2m)}^{a/2}{(w_{a}(x))}^{-1/2},
\end{equation}
which follows from the three-term recursion and a sum rule for
$L_{j}^{(a)}(x)$ \cite{szego}.


\subsection{Level-density and 2-point correlation for associated Laguerre symplectic ensemble}

In this subsection, we calculate the level density of associated
Laguerre symplectic ensemble and show that in the bulk of the
spectrum, the scaled or ``unfolded'' \cite{mehta} kernel function
$(S^{(4)}_{2N}(x,y)/S^{(4)}_{2N}(y,y))$ is stationary and universal.

To study the kernel function (\ref{rgcd4}), we need to calculate $R^{(4)}_{2N-1,2N+1}$,
$R^{(4)}_{2N-1,2N}$, $R^{(4)}_{2N-2,2N}$ and $P^{(4)}_{2N-1,2N}$.
For this, we use (\ref{p4Jquatelem}) and (\ref{r4Jquatelem}) for
associated Laguerre weight:
\begin{eqnarray}
&& x\frac{d}{dx}\phi^{(4)}_{2m+1}(x)=\sum^{2m+2}_{j=2m-2}P^{(4)}_{2m+1,j}\phi^{(4)}_{j}(x),\\
&&
x^2\frac{d}{dx}\phi^{(4)}_{2m+1}(x)=\sum^{2m+3}_{j=2m-2}R^{(4)}_{2m+1,j}\phi^{(4)}_{j}(x),\qquad
x^2\frac{d}{dx}\phi^{(4)}_{2m}(x)=\sum^{2m+2}_{j=2m-2}R^{(4)}_{2m,j}\phi^{(4)}_{j}(x).
\end{eqnarray}
From which, using properties of OP, we get
\begin{eqnarray}
&& P^{(4)}_{2m+1,2m+2}=\sqrt{\frac{g^{(4)}_{2m+2}}{g^{(4)}_{2m}}}(m+1),\qquad
 R^{(4)}_{2m+1,2m+3}=-\sqrt{\frac{g^{(4)}_{2m+2}}{g^{(4)}_{2m}}}(m+1)(2m+3),\\
&&
R^{(4)}_{2m,2m+2}=-\sqrt{\frac{g^{(4)}_{2m+2}}{g^{(4)}_{2m}}}(m+1)(2m+1),\qquad
R^{(4)}_{2m+1,2m+2}=\sqrt{\frac{g^{(4)}_{2m+2}}{g^{(4)}_{2m}}}(m+1)(4m+a+4).
\end{eqnarray}
For $m=N-1$, large $N$, we have $(g^{(4)}_{2N}/g^{(4)}_{2N-2})\simeq
1$. Finally for large $N$, the recursion coefficients are given by
\begin{eqnarray}
\nonumber
&& P^{(4)}_{2N-1,2N}\sim N+O(1),\\
\label{approxPR4lag} && R^{(4)}_{2N-1,2N+1}\sim -2N^2+O(N),\qquad
R^{(4)}_{2N-2,2N}\sim -2N^2+O(N),\qquad R^{(4)}_{2N-1,2N}\sim
4N^2+O(N).
\end{eqnarray}
Also, since $g^{(4)}_{2N-2}/g^{(4)}_{2N}\simeq 1$, we can write

\begin{eqnarray}
\nonumber \phi^{(4)}_{2N}(x)\phi^{(4)}_{2N-1}(y) &=&
\frac{1}{\sqrt{{g^{(4)}_{2N-2}g^{(4)}_{2N}}}}
\sqrt{w(x)w(y)}\pi_{2N}(x)\pi_{2N-1}(y)\\
\nonumber
&\simeq & \frac{1}{g^{(4)}_{2N}}\sqrt{w(x)w(y)}\pi_{2N}(x)\pi_{2N-1}(y)\\
\nonumber
&=& - \frac{1}{h^{a}_{2N}}\sqrt{w(x)w(y)}\pi_{2N}(x)\pi_{2N-1}(y)\\
\label{negetiveg} &\simeq & -
{(2N)}^{-a}\sqrt{w(x)w(y)}\pi_{2N}(x)\pi_{2N-1}(y).
\end{eqnarray}
Finally using the asymptotic results (\ref{pioddasylag4}),
(\ref{pievenasylag4}), (\ref{pievenasylag4delta}), defined in the
$\theta$ interval $[\epsilon,\pi/2-\epsilon N^{-\frac{1}{2}}] $, and
the large $m$ results (\ref{approxPR4lag}) and (\ref{negetiveg}) in
the GCD formula (\ref{rgcd4}), we get for

\begin{eqnarray}
\nonumber
y=(8N+2a+4)\cos^2\theta,\qquad && x=y+\Delta y,\\
 \nonumber
y(y-x)S^{(4)}_{2N}(x,y) &=&
R^{(4)}_{2N-2,2N}\left[\phi^{(4)}_{2N}(x)\phi^{(4)}_{2N-1}(y)
-\phi^{(4)}_{2N}(y)\phi^{(4)}_{2N-1}(x)\right]\\
\nonumber && +
R^{(4)}_{2N-1,2N+1}\left[\phi^{(4)}_{2N-2}(x)\phi^{(4)}_{2N+1}(y)
-\phi^{(4)}_{2N-2}(y)\phi^{(4)}_{2N+1}(x)\right]\\
\nonumber
&& +(R^{(4)}_{2N-1,2N}-
xP^{(4)}_{2N-1,2N})\left[\phi^{(4)}_{2N-2}(x)\phi^{(4)}_{2N}(y)
-\phi^{(4)}_{2N-2}(y)\phi^{(4)}_{2N}(x)\right] \\
\nonumber &=&
-\frac{N}{2\pi\sin^{2}\theta}\left\{\sin[f_{2N-2}(\theta)-f_{2N}(\theta+\Delta\theta)]
+\sin[f_{2N}(\theta)-f_{2N-2}(\theta+\Delta\theta)]\right\}\\
\nonumber && + \frac{N\cos
2\theta}{4\pi\cos\theta\sin^{3}\theta}\left\{\cos[f_{2N-2}(\theta+\Delta\theta)]\cos[(f_{2N}(\theta)]
-\cos[f_{2N-2}(\theta)]\cos[f_{2N}(\theta+\Delta\theta)]]\right\}\\
\nonumber &=& \left[\frac{N\cos 4\theta}{\pi\sin^2\theta}
-\frac{N\sin 4\theta\cos
2\theta}{4\pi\cos\theta\sin^3\theta}\right]\sin\left(\frac{\partial
f_{2N}}{\partial\theta}\Delta\theta\right)\\
&=& 8N\cos^2\theta \frac{\left[\cos 4\theta-\cos^2 2\theta\right]}{
{2\pi\sin^2 2\theta}}\sin\left(\frac{\partial
f_{2N}}{\partial\theta}\Delta\theta\right),
\end{eqnarray}
which gives
\begin{eqnarray}
\nonumber \Delta y S^{(4)}_{2N}(x,y) = -\frac{\sin
(8N\sin^2\theta\Delta\theta)}{2\pi}.
\end{eqnarray}
Thus we get
\begin{eqnarray}
S^{(4)}_{2N}(x,y) &=&
\frac{\sin\left(\frac{1}{2}y^{-\frac{1}{2}}\sqrt{8N-y}\Delta
y\right)}{2\pi\Delta y},\qquad 0<y< 8N.
\end{eqnarray}
With $\Delta y\rightarrow 0$, we get the level density
\begin{eqnarray}
S^{(4)}_{2N}(y,y) &=& \frac{1}{4\pi}\sqrt{\frac{8N-y}{y}},\qquad
0<y< 8N.
\end{eqnarray}
Finally, in the bulk of the spectrum, we get the ``universal''
sine-kernel (\ref{sin4}).

\section{Conclusion}

In \cite{ghosh} and \cite{ghoshpandey}, the authors prove
``universality'' for the entire Jacobi ensembles of random matrices
using SOP, which were written in terms of OP. The
asymptotic properties of OP were used to obtain
asymptotic formul\ae\ for SOP. Finally, the summation in
$S_{2N}^{\bt}(x,y)$ was replaced by an integral for large $N$.
However, using asymptotic results of the SOP to calculate the kernel
function, before completing the sum lacks mathematical rigor.

In this paper, we have shown that SOP
$\Phi_{n}^{\bt}(x)$ and $\Psi_{n}^{\bt}(x)$ corresponding to
classical weight satisfy three-term recursion relations in the
$2\times 2$ quaternion space. Using this, we obtain the kernel
functions $S_{2N}^{\bt}(x,y)$, $\beta=1,4$, for the entire family of
finite dimensional Jacobi ensembles of random  $2N\times 2N$
matrices. As $N\rightarrow\infty$, we use the asymptotic results of
the SOP in the range $[x_{1}+\epsilon,x_{2}-\epsilon]$, over which
they are defined, to prove that in the bulk of the spectrum, the
correlation functions are universal.

Here, we would like to mention that our GCD results are valid in the
entire complex plane. Hence to  study statistical properties of the
eigenvalues of orthogonal and symplectic ensembles away from the
bulk, we need to use Plancherel-Rotach type formula for these
polynomials defined outside the range
$[x_{1}+\epsilon,x_{2}-\epsilon]$, something which has already been
done for the unitary ensembles. We wish to come back to this in a
later publication.

We would like to emphasize that the key step in deriving the
asymptotic results of SOP is to obtain and solve  finite term recursion relations between
SOP and OP. Recently, this technique has been used
 to obtain bulk asymptotics of SOP corresponding to quartic double well potential \cite{ghosh5}. Till
date, this seems the easier method to study the asymptotic behavior
of SOP rather than solving the $2d\times 2d$ Riemann
Hilbert problem \cite{ghosh4,pierce}.

Finally, a word on duality. Results in section 3 makes us wonder if
the two families of SOP are really different or there exists a
simple mapping between them. We have seen the existence of a simple
relation (\ref{dualityrelation}) between SOP
arising in the study of the associated Laguerre ensembles. The other
ensembles do show similar pattern, although we are unable to come
out with a general formula. A deeper theoretical understanding is
needed to obtain the mapping between these two families of
SOP.

\end{document}